\documentclass[a4paper,fleqn,usenatbib]{mnras}

\usepackage{newtxtext,newtxmath}

\usepackage[T1]{fontenc}
\usepackage{ae,aecompl}
\usepackage{ulem}

\usepackage{graphicx}	
\usepackage{amsmath}	
\usepackage{amssymb}	

\newcommand{\pc}{{\rm \,\pc}}

\title[Jet Feedback in Star Formation]{The Effects of Protostellar Jet Feedback on Turbulent Collapse}

\author[Murray, Goyal, \& Chang]{Daniel Murray$^1$\thanks{dwmurray@uwm.edu}, Shivam Goyal$^2$, and Philip Chang$^1$
\\
  $^{1}$Department of Physics, University of
Wisconsin-Milwaukee, 3135 North Maryland Ave., Milwaukee, WI 53211, USA\\
  $^{2}$Department of Physics and Astronomy, Dartmouth College, 6127 Wilder Laboratory, Hanover, NH 03755-3528,  USA\\
}

\date{Accepted XXX. Received YYY; in original form ZZZ}

\pubyear{2017}

\begin{document}
\label{firstpage}
\pagerange{\pageref{firstpage}--\pageref{lastpage}}
\maketitle

\begin{abstract}
We present results of hydrodynamic simulations of massive star forming regions with and without protostellar jets.
We show that jets change the normalization of the stellar mass accretion rate, but do not strongly affect the dynamics of star formation.
In particular, $M_*(t) \propto f^2 (t-t_*)^2$ where $f = 1 - f_{\rm jet}$ is the fraction of mass accreted onto the protostar, $f_{\rm jet}$ is the fraction ejected by the jet, and $(t-t_*)^2$ is the time elapsed since the formation of the first star.
The star formation efficiency is nonlinear in time.
We find that jets have only a small effect (of order 25\%) on the accretion rate onto the protostellar disk (the ``raw'' accretion rate).
We show that the small scale structure -- the radial density, velocity, and mass accretion profiles are very similar in the jet and no-jet cases.
Finally, we show that the inclusion of jets does drive turbulence but only on small (parsec) scales.
\end{abstract}




\section{Introduction}

On galactic scales, star formation is observed to be slow as implied by the Kennicutt-Schmidt law \citep{1998ApJ...498..541K,2008AJ....136.2782L}:
\begin{equation}\label{eq:kennicutt}
 \dot{M}_* = \epsilon \frac{M_g}{\tau_{\rm dyn}}
\end{equation}
where $\epsilon = 1-2\%$ of the mass of the gas, $M_g$ is converted to stars, $M_*$, per dynamical time, $\tau_{\rm dyn}$. On smaller scales, i.e., the scales of giant molecular clouds (GMCs), observations show a large scatter in the efficiency, $\epsilon$, in both GMCs \citep{2016ApJ...833..229L} and smaller clouds \citep{2014ApJ...782..114E}. 

A number of explanations for this low star formation rate and scatter in the efficiency, on either galactic or GMC scales, have been put forth. 
On large scales, the leading candidate is stellar feedback, e.g. \citet{1986ApJ...303...39D}, in which supernovae limit the amount of dense gas. 
On small scales, these include turbulent pressure support \citep{1992ApJ...396..631M} and support from magnetic fields \citep{1966MNRAS.132..359S,1976ApJ...207..141M}. 
Numerical experiments investigating turbulence and magnetic fields suggest that, while magnetic support found in MHD simulations can slow the rate of star formation compared to hydrodynamics simulations, neither turbulence nor magnetic support is sufficient to limit the small scale star formation rate to 1-2\% per free fall time
\citep{2010ApJ...709...27W,2011MNRAS.410L...8C,2011ApJ...730...40P,2012ApJ...754...71K,2014MNRAS.439.3420M,2015ApJ...808...48B,2017ApJ...838...40M}.

\citet{2015ApJ...800...49L} showed that, in simulations with no feedback, the star formation efficiency on
parsec scales is not constant in time.
This is in contrast to previous work, which had implicitly assumed that the star formation rate on small scales was constant.
In particular, many authors have assumed that the star formation rate in their simulations of a GMC (or smaller cloud or part of a cloud) was given by equation (\ref{eq:kennicutt}), where $\epsilon$ was assumed to be constant.
\citet{2015ApJ...800...49L} showed that $\epsilon \propto t$, which implies that $M_* \propto t^2$, where $M_*$ is the total stellar mass.
Motivated by this, \citet[][hereafter MC15]{2015ApJ...804...44M} developed a new 1-D model of spherical collapse that treats the turbulent velocity, $v_T$, as a dynamical variable.
Following earlier work \citep{1977ApJ...214..488S,1992ApJ...396..631M,1997ApJ...476..750M}, MC15 reduced the fluid equations by assuming spherical symmetry, but did not assume a fixed equation of state.
Early work \citep{1977ApJ...214..488S,1992ApJ...396..631M,1997ApJ...476..750M} assumed that the pressure, which appears in the momentum equation, was given by $P = \rho c_s^2$ with $c_S = \textrm{const}$ or $P = \rho v_{\rm T}^2$ with $v_{\rm T} = v_{\rm T}(r) \propto r^{\kappa_r}$, i.e., a prescribed function of $r$, and $\kappa_r$ is typically chosen to be 1/2 to reflect Larson's law \citep{1981MNRAS.194..809L}. 
In contrast,  MC15 used the results of \citet{2012ApJ...750L..31R} on compressible turbulence including both decay and compression, treating the turbulent velocity as a dynamical variable.  In essence, MC15 includes an extra equation for the evolution of the energy.
In doing so, MC15 found that the $t^2$ power law arises naturally because the small scale density and velocity are set by the gravity of the protostar, and made predictions for the velocity and mass accretion 
profiles that were later verified numerically by \citet{2017MNRAS.465.1316M}.

In any case, the issue of the slow rate of large scale star formation remains.
One possibility is that stellar feedback acting on large (galactic disk) scales controls the rate of star formation on those scales.
Cosmological simulations including both radiative and supernova feedback can reproduce Kennicutt's observational results, e.g., \cite{2011MNRAS.417..950H,2013ApJ...770...25A,2014MNRAS.445..581H}. If these simulations are to be believed, the physics of stellar feedback from protostellar jets, which are not included, may not control the global galactic rate of star formation.

However, jets may control the rate of small (parsec) or medium (GMC) scale star formation and may power the observed turbulence in molecular clouds \citep{2007ApJ...659.1394M}.
As a result, a number of groups have recently studied the effects of protostellar jets and outflows on star formation \citep{2010ApJ...709...27W,2014MNRAS.439.3420M,2015MNRAS.450.4035F}.
\citet{2007ApJ...662..395N} and \citet{2011ApJ...740...36N} found that the effect of protostellar outflows was important for driving turbulence.  \citet{2011ApJ...740..107C} and \citet{2012ApJ...747...22H} found that protostellar outflows enhanced the effectiveness of radiative feedback.  In addition, \citet{2015MNRAS.450.4035F} found that a combination of turbulence, jets, and magnetism is able to reproduced the observed low efficiency (to within a factor of four) of star formation in contrast to galactic scale simulations that rely on radiative and supernovae feedback.

The results of \citet{2015ApJ...800...49L}, \citet{2015ApJ...804...44M}, and \citet{2017MNRAS.465.1316M} demonstrated that the small scale dynamics are intimately linked to the star formation rate. The question we address in this paper is, do protostellar jets affect the small scale physics of accretion in a turbulent medium?  In particular, does the stellar mass still increase as $t^2$ with the inclusion of protostellar jet feedback?  The answer as we will argue below is no and yes, respectively.

The paper is organized as follows.
In \S~\ref{sec:implementation}, we describe the numerical implementation of the jet outflow (\S~\ref{sec:jet feedback}).
In \S~\ref{sec:protostar evolution} we describe our protostar evolutionary model, which is based on the one-zone models of \citet{2000ApJ...534..976N} and \citet{2009ApJ...703..131O}.
We then present our results in \S~\ref{sec:turbulent star formation}, describing the large (pc) scale jet effects (qualitatively) in \S~\ref{sec:large scale}.
In \S~\ref{subsec:SFR} we show that the star formation rate is reduced by the ejection of mass in the jet, but the jets do not, surprisingly, significantly change accreting gas properties (in a spherically averaged sense) in \S ~\ref{sec:run_density} - \ref{subsec:mdotstars}.
We discuss the effects of the jet on the gas in terms of momentum depositions in  \S~ \ref{subsec:jet_mv} and the driving of turbulence in \S~\ref{sec:jets_drive_small}.
We discuss our results in \S~\ref{sec:discussion} and give our conclusions in \S~\ref{sec:conclusions}.

\section{Numerical Implementation}\label{sec:implementation}

\subsection{Jet Feedback Prescription}\label{sec:jet feedback}

We have implemented a model of jet feedback in the adaptive mesh refinement (AMR) code, RAMSES
 \citep{2002A&A...385..337T}.  Ramses is a mature AMR code that include self-gravity, sink particle formation \citep{2010MNRAS.409..985D,2014MNRAS.445.4015B}, and radiative transfer \citep{2013MNRAS.436.2188R,2015MNRAS.449.4380R}.  RAMSES has been used in a number of problems including cosmological structure formation (e.g., see for instance \citealt{2016MNRAS.463.1462O}), star cluster formation \citep{2017MNRAS.472.4155G}, 
colliding winds \citep{2011MNRAS.418.2618L,2017MNRAS.468.2655L}, and relativistic astrophysics \citep{2013A&A...560A..79L,2017arXiv170204362L}. More recently, we have added a turbulent stirring module and have used this for star formation simulations in a turbulent gas without feedback \citep{2017MNRAS.465.1316M}.

Here we build on our previous work \citep{2017MNRAS.465.1316M} by adding jets to the sink particle implemention in RAMSES \citep{2010MNRAS.409..985D,2014MNRAS.445.4015B}.   A jet is launched once a star particle has a mass in excess of $0.01 M_\odot$ at which point it also has a well defined spin angular momentum.
When the sink particle accretes gas from nearby cells, a fraction, $f_{\rm jet}$, with a fiducial value $f_{\rm jet} = 0.3$, of this gas is launched along the spin axis of the protostar.  
Numerically, this involves the injection of mass and momentum into nearby cells, while subtracting the ejected mass from the sink particle.
The injection region consists of a bi-cone with an opening angle of $0.3$ radians ($\approx 17$ degrees) about the spin axis of the sink particle
and a radial extent between 4 and 8 cells (at the highest refinement level) away from the sink particle.
In particular, we set per cell in the injection regions:
\begin{eqnarray}
\dot{\rho} &=& \sum_i^{n_{\rm sink}} f_{\rm jet}\mathcal{R}(r)\mathcal{T}(\theta), \\
\dot{\bf \pi} &=& \sum_i^{n_{\rm sink}}\dot{\rho}_i {\bf v}_{\rm jet},
\end{eqnarray}
where $\mathcal{T}$ is the angular distribution, $\mathcal{R}$ is the radial distribution, $r$ is the radial distance of the cell centre from the $i$th sink particle.

For the angular distribution, $\mathcal{T}$, we assume a Gaussian jet (in angle) about the spin axis of the sink particle:
\begin{equation}
\Theta(\theta) \propto \exp\left(-\frac{\theta^2}{\theta_0^2}\right) \approx  \exp\left(\frac{1-\cos\theta}{\cos\theta_0 - 1}\right),
\label{eq:theta_approx}
\end{equation}
where we choose $\theta_0 \ll 1$. 
In practice we use the approximation in equation (\ref{eq:theta_approx}) because $\cos\theta = \hat{l}_{\rm sink} \cdot \hat{r}_{\rm sink} $, where ${\bf l}_{\rm sink}$ is the angular momentum vector of the sink particle, and $\hat{r}_{\rm sink}$ is the direction from the ith sink particle to the cell in question, which simplifies the numerical implementation and computation, i.e., we avoid an $\cos^{-1}$ computation.

For the radial distribution, $\mathcal{R}$, we set the deposition of mass and momentum to be constant between $r_{\rm jet, in} = 4$ and $r_{\rm jet, out} = 8$ grid cells from the sink particle.
To ensure that this region is resolved, we force the region $r<r_{\rm jet,out}$ around a sink particle to be  refined to the maximum level.

To further ensure that the injection of mass and momentum per grid cell is as accurate as possible, we have integrated the injection of mass and momentum over the entire cell rather than evaluate the value at the cell center.  This is necessary as the injection region depends on the grid cell size and is not necessarily a physical region.
Here, we discretized the cell into $n_{\rm sg}^3$ sub-cells, where $n_{\rm sg}$ is typically set to 3. This gives 27 evaluations of the mass and momentum injection per cell.
The central value of the injected mass and momentum was taken for each of the sub-cells, and subsequently averaged to find the effective mass and momentum deposition over the entire cell.
This algorithm was used because it gave more accuracy in the jet feedback implementation.
Our method is similar to the methodology in \citet{2011ApJ...740..107C} and \citet{2014MNRAS.439.3420M} to conserve mass and momentum where they compute the normalization of the jet kernel -- effectively $\Theta(\theta)$ -- numerically.
Similarly, \citet{2014ApJ...790..128F} iterates over the deposition of jet mass and momentum per star particle to conserve mass and momentum exactly.

We use the protostellar mass and radius to set the jet velocity.
In particular, we set
\begin{equation}\label{eq:vjet}
v_{\rm jet} = f_{K} \sqrt{\frac{GM_*}{r_*}},
\end{equation}
where $f_{K}$ is the fractional percentage of the Keplerian velocity the jet reaches asymptotically, and $r_*$ is the radius of the accreting protostar.
We take $f_{K} = 1/3$, following \citet{2014MNRAS.439.3420M}.
To set $r_*$, we use two methods.
First, following \citet{2014ApJ...790..128F} we fix the protostellar radius at $r_* = 10\,R_{\odot}$.
We use this only in the next section for our test problem.
Second, we implement a one-zone protostellar model described by \citet[][hereafter known as O09]{2009ApJ...703..131O} and utilized in \citet{2014MNRAS.439.3420M}.
This latter model is used in our turbulent star formation simulations below.

\subsubsection{Test Problem}

To test the physics of the jet and our numerical implementation, we consider the following test problem.
We simulate a periodic box with a length of 0.2 pc on each side and a fixed background density $\rho_0 = 3\times 10^{-20}\,{\rm g\,cm^{-3}}$ ($n_0 = 10^4$ for H$_2$) of isothermal gas with a temperature of $\approx 17$ K, which corresponds to a sound speed of $0.265\,{\rm km\,s}^{-1}$.
This setup is similar to the turbulent star formation simulations performed below and in \citet{2015ApJ...800...49L} and \citet{2017MNRAS.465.1316M}, but with a larger density and a smaller box.
In the centre, we place a region of higher density, i.e., a clump, with a Gaussian density distribution, $\rho(r) = \rho_m\exp(-r^2/r_0^2)$,  with maximum density of $\rho_m=3\times 10^{-16}\,{\rm g\,cm^{-3}}$ and a characteristic radius $r_0=0.017$ pc.
The total mass of this dense region is approximately $M = 100\,M_{\odot}$.
The entire clump is set to solid body rotation about the z-axis with an angular frequency corresponding to 10\% of the breakup velocity of the clump.
The sink particle that forms also has a spin along the z-axis.

\begin{figure*}
	\includegraphics[width=2.2\columnwidth]{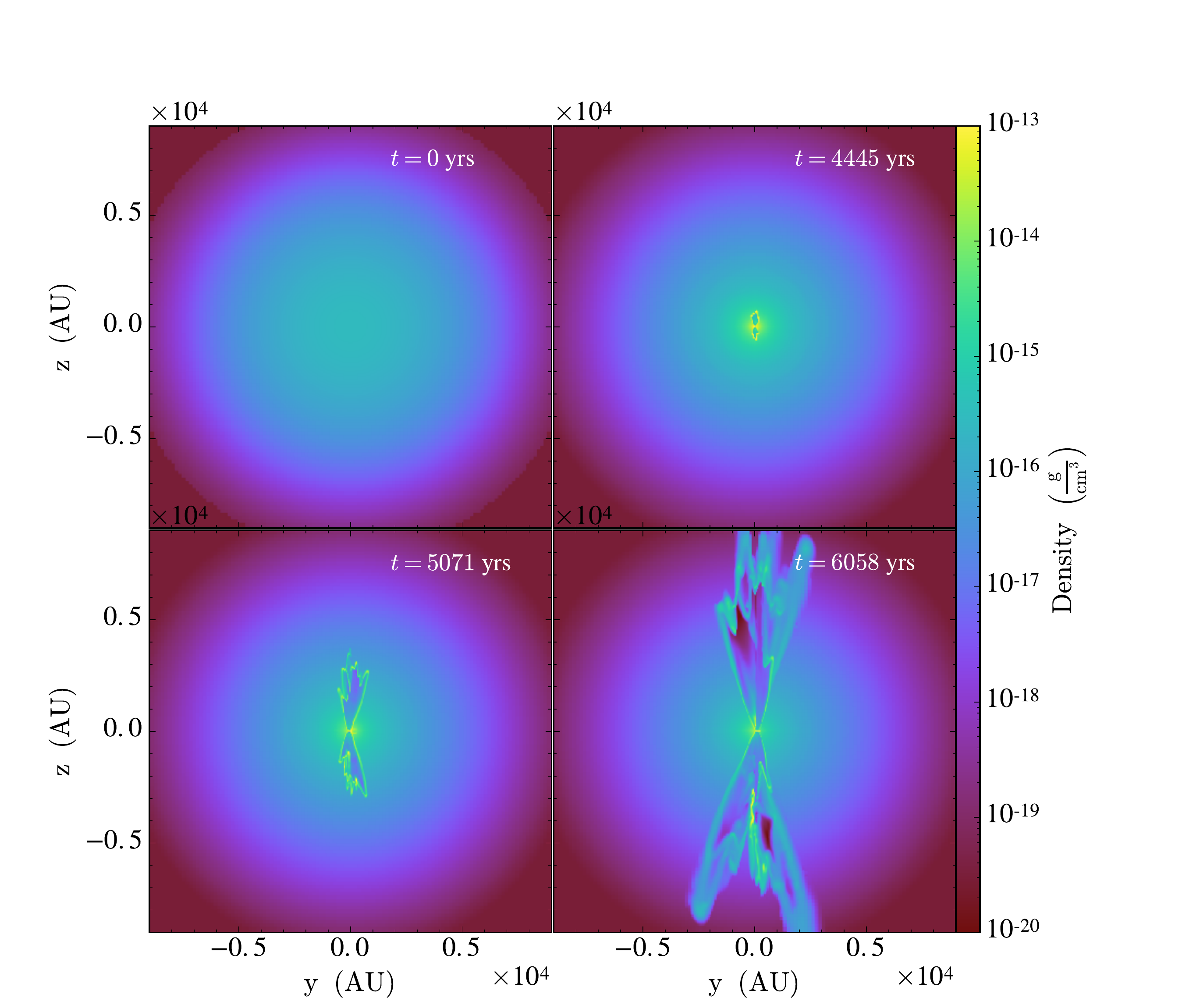}
    \caption{Sequential plots of density structure perpendicular to the disk plane through the centre of the sink particle. Columns from left to right show different times, from $t=0$ to $t=4820$ yrs, showing the effect of jet feedback on the test problem.}
    \label{fig:test_jet}
\end{figure*}

In Figure \ref{fig:test_jet}, we plot 4 snapshots of the collapsing clump.
At $t=0$ (upper left panel), the simulation is initialized as described above.
This clump begins to collapse, reaching a peak density of $10^{-13}\,{\rm g\,cm}^{-3}$ and forms a sink particle with its spin axis oriented along the z-axis at $t= 4445$ yrs (upper right panel).
A jet develops and launches $f_{\rm jet} = 1/3$ of the accreted mass along that axis.
This jet grows and extends across the entire plotted region by $t=6058$ yrs (lower left and right panels).


\subsection{Protostar Evolution Prescription}\label{sec:protostar evolution}

In our jet model, the jet velocity is scaled to the Keplerian velocity at the surface of the protostar.
Hence, both the mass and radius of the protostar is required.  As discussed above, we can either fix the radius of the protostar as \citet{2014ApJ...790..128F} have done or implement a one-zone protostellar model (O09).

We now describe the latter case. Our treatment follows that of \citet{2014MNRAS.439.3420M} and O09. 
This 1-zone model, which is originally due to \citet{2000ApJ...534..976N}, describes the evolution of the radius of the protostar due to accretion, cooling, gravitational contraction, and nuclear burning.



Following the formation of a sink particle, we step through the following algorithm. 
\begin{enumerate}
 \item If the sink particle has $M_*$ below $0.01 M_{\odot}$ (the ``pre-collapse'' state) no jet is launched.
 \item Once a star particle exceeds $0.01 M_{\odot}$ it is assigned a radius and polytropic index based on the mass accretion rate (their equations (B1)-(B3) in O09).
The deuterium mass is scaled to the mass of the sink particle times the cosmological abundance of deuterium.
The protostar's state is set to be ``no burning,'' because its core temperature is below that needed to burn deuterium.
 \item We evolve the protostar in accordance to O09's equation (B4), which is a discretized version of equation (8) of \citet{2000ApJ...534..976N}.
We reproduce it here for convenience:
 \begin{eqnarray}
\label{eq:deltaR}
\Delta r_* &=& 2 \frac{\Delta M_*}{M_*} \left( 1 - \frac{1 - f_k}{\alpha_g \beta} + \frac{1}{2} \frac{d\, \rm{log} \beta}{d\, \rm{log} M_*} \right) r_* - 2 \frac{\Delta t}{\alpha_g \beta} \nonumber\\
&&\times \left( \frac{r_*}{G M_*^2} \right) \left( L_{int} + L_{ion} - L_D \right) r,
\end{eqnarray}
where $\alpha_g = 3/(5 - n)$ describes the gravitational binding energy of a polytrope, and $\beta(n, M_*, r_*)$ is the ratio of gas pressure to the total pressure for a polytrope of index $n$ and (proto)stellar mass $M_*$ and radius $r_*$ as defined above.
$f_k$ is the fraction of kinetic energy of the infalling material that is radiated away (typically $f_k = 0.5$), $L_{\rm int}$ is the luminosity leaving the stellar interior, $L_{ion}$ is the luminosity required to dissociate and ionize all infalling material, and $L_D$ is the luminosity which is supplied by deuterium burning.
We precompute $\beta$ as a function of $n$, $M_*$, and $r_*$ and linearly interpolate with that table for specific values. The luminosities, $L_{\rm int}$, $L_{\rm ion}$, and $L_D$ are determined from equations (B6)-(B9) of O09 respectively.
\item Once the central temperature, reaches $T_c \geq 1.5\times 10^6$ K, the protostar advances to a ``core burning at fixed temperature'' state.  Following O09, we also reset $n=1.5$.  We procede to burn deuterium at a rate to maintain a fixed temperature in accordance with O09 (their equation (B8)).
\item Once the deuterium mass in the star drops to zero, the protostar switches to ``core burning at variable core temperature''; deuterium is burned as rapidly as it is accreted (O09, their equation (B9)).
\item At this point the star can take two paths.
If the radius of the star falls below $r_{\rm ms}$, where $r_{\rm ms}$ is the radius of the main sequence star of the same mass (from \citet{1967CaJPh..45.3429E},) we set $r_*=r_{\rm ms}$ and the state of the star is set to ``main sequence''.
However, if $L_D > 0.33 L_{\rm ms}$, where $L_{\rm ms}$ is the main sequence luminosity, we set the protostellar state to be ``deuterium shell burning'', $n\rightarrow 3$, and the radius is expanded by a factor of 2.1 to mimic swelling due the formation of a radiative zone (O09).
Subsequently the star then shrinks down to the main sequence.
\end{enumerate}
This model provides a protostellar radius for each star particle, which we then use in equation (\ref{eq:vjet}) to determine the jet velocity.
Though we wait until the initial mass of our protostar is $0.01 M_\odot$, our choice for the threshold mass does not greatly alter the evolution of the system.
As stated in equation (\ref{eq:vjet}): $v_{\rm jet} \propto M_*^{1/2}$.
The mass expelled by the jets is determined by: $m_{\rm jet} = f_{\rm jet} \delta M$, where  $\delta M$ is the total mass that would be accreted by the sink particle in the absence of jet feedback. 
It can be shown that $\dot{m}_{\rm jet} \propto m^{1/2}$.
The momentum of the jet is simply: $p_{\rm jet} = m_{\rm jet} * v_{\rm jet}$; 
thus, the time rate of change in momentum due to the mass is: $\dot{p}_{\rm jet} \approx \dot{m}_{\rm jet} * v_{\rm jet}$.
Writing this expression solely as a function of mass: $\dot{p}_{\rm jet} \propto m^{1/2} * m^{1/2} = m_*$. 
Therefore the total integrated momentum: $p_{\rm jet} \approx \frac{2}{3} m^{3/2}$.
So, if we do a comparison between the momentum of the jet from our threshold mass of $0.01$ to $0.1 M_\odot$ vs $0.1$ to $1.0 M_\odot$, we see that changing our threshold mass from $0.01 M_\odot$ to $0.1 M_\odot$ affects our result by $\approx 0.1 \%$. 

\section{Jet Feedback in Turbulent Star Formation}\label{sec:turbulent star formation}

We use our protostellar star and jet model in RAMSES \citep{2002A&A...385..337T} to model self-gravitating, hydrodynamic turbulence in isothermal gas with three-dimensional (3D), periodic grids.
We use eight levels of refinement on a root grid of $128^3$, giving an effective resolution of $32K^3$.
We also performed runs at $8K^3$ and $16K^3$ to confirm convergence.
We start with a box with the physical length set to $L = 16$ pc using periodic boundary conditions with an initial mass density of $\rho = 3\times10^{-22}\,{\rm g\,cm}^{-3}$ (number density $n \approx 100\,{\rm cm}^{-3}$ for molecular hydrogen), corresponding to a mean free-fall time $\bar\tau_{\rm ff}\approx3.8\,{\rm Myrs}$.
This gives a total mass in the box $M\approx 18,000\,M_\odot$.
We fix the sound speed to be $c_s = 0.264\,{\rm km\,s^{-1}}$, which for pure molecular hydrogen corresponds to an ambient temperature of $T \approx 17 {\rm K}$.

To initialize our simulations, we drive turbulence by applying a large scale
($1 \le kL \le 2$) fixed solenoidal acceleration field as a momentum
source term. 
We apply this field in the absence of gravity and star particle formation for about 3 dynamical times until a statistical steady state is reached.
At this point, the density is no longer uniform, instead it has a log-normal distribution.

After this statistical steady state is reached, we turn on self-gravity and star particle formation for our star formation experiments. 
We refine collapsing regions using a modified Truelove criterion, where
$\lambda_J \le N_J \Delta x$, where $\Delta x$ is the cell length and $N_J$ is the number of cells per Jeans length, $\lambda_J$ \citep{1997ApJ...489L.179T}.
This corresponds to a condition on the density
\begin{eqnarray}
\frac{\rho}{\rho_0} &=& 11 \cdot 4^l
\left( \frac{N_{\rm root}}{128} \right)^2 \left( \frac{N_{J}}{8} \right)^{-2}
\left( \frac{16 \rm{pc}}{L} \right)^2
\left( \frac{c_s}{0.265\, {\rm km\, s}^{-1}} \right)^2\nonumber\\
&&\times\left( \frac{3 \times10^{-22} {\rm g\, cm}^{-3}}{\rho_0} \right)
\label{eq:refinement_criteria}
\end{eqnarray}
where $l$ is the refinement level, with $l = 0$ corresponding to the root grid.
We use $N_J=8$ to ensure that the Jeans length is resolved by at least 8 cells.
When this density condition is met the local grid is refined by a factor of 2,
provided that the maximum refinement level has not been reached.

When the Truelove criterion 
is exceeded by a factor of three at the highest refinement level, the excess mass is either accreted onto a nearby sink particle, used to create a new sink particle, or left alone.
If the distance to the nearest sink particle is less than $2$ grid cells, then the material is accreted onto that sink particle.
On the other hand, if the gas is contracting (local divergence of velocity is negative), a local potential minimum, and sufficiently far away from other sink particles, then the excess mass is used to produce a sink particle.
Finally, if these additional checks are not satisfied, then the gas is left alone.
We should note that this star formation criteria differs from our previous work in \citet{2015ApJ...800...49L} and \citet{2017MNRAS.465.1316M}, which did not include these additional checks.
We have found that jet feedback produces small pockets of high density as a result of shocks from the jet.  This is why we have included these new checks in this work.
To make a fair comparison between the feedback and no-feedback cases, we run the same numerical experiments both with and without jets.

Whether a star particle is newly formed from the collapsing gas or accreting the surrounding gas, it keeps track of its angular momentum.
As a result each star particle is endowed with an angular momentum vector from which we can apply our jet feedback prescription.
In addition, we have also modified each star particle to track its protostellar state which changes based on the conditions laid out above.

\subsection{Parsec scale effects of the jets}\label{sec:large scale}

\begin{figure*}
	\includegraphics[width=0.8\textwidth]{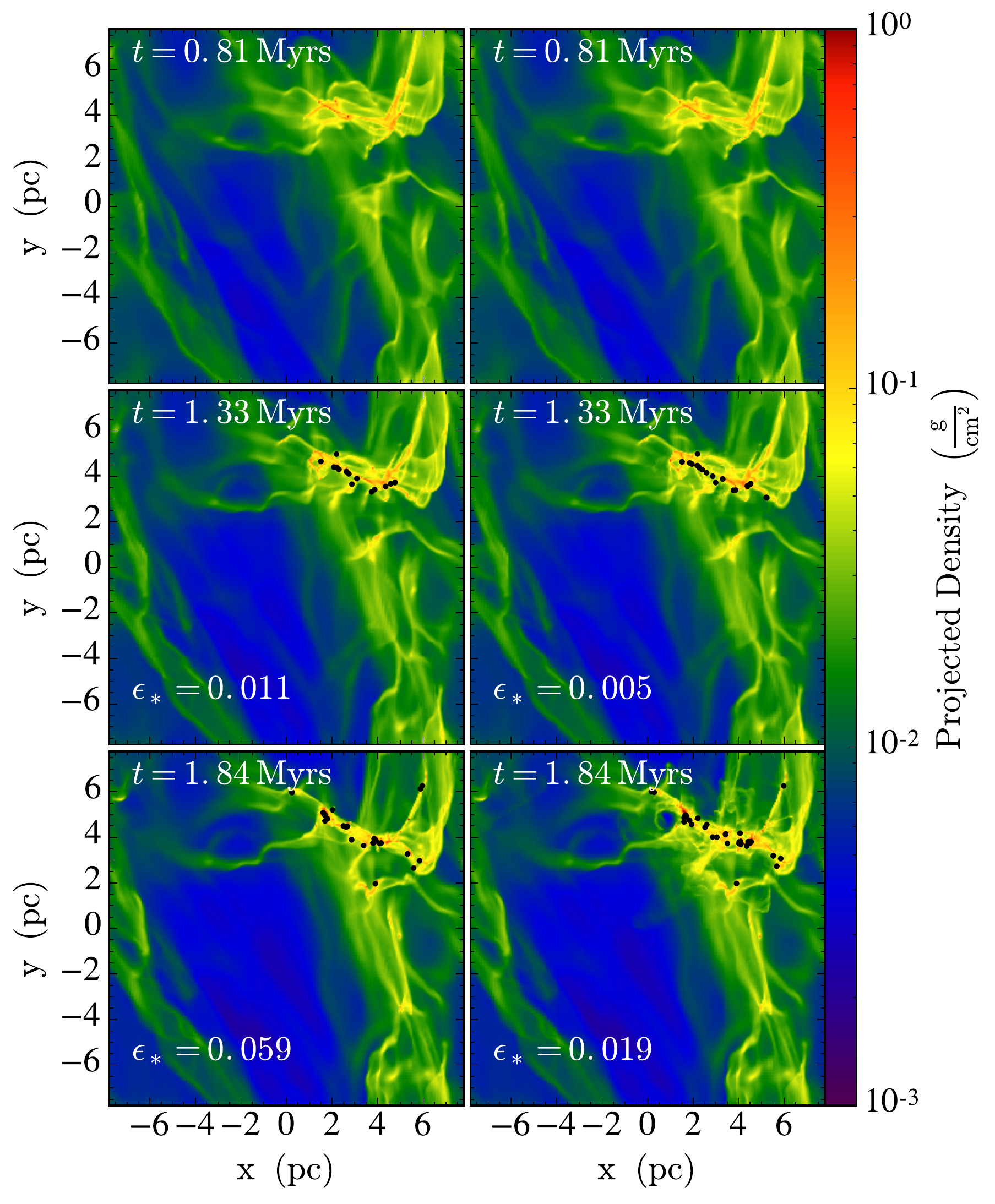}
    \caption{Sequential projections of the density structure of the entire simulation domain along the z-axis for simulations with (right column) and without (left column) jet feedback.  From top to bottom, we show the plots at $t=0.8$ Myrs (top), which is right after the first stars form to $t=1.33$ Myrs (middle) and $t=1.84$ Myrs (bottom).  The SFE advances from $\epsilon_* = 0$ (right top) to $0.019$ (right bottom) for the case with jet feedback and from $\epsilon_* = 0$ (left top) to $0.06$ (left bottom) in the case without jet feedback. The black dots represent locations where one or more sink particles have been created. We state one or more sink particles, as the full box projections are zoomed out such that small clusters of two or three sinks have been plotted under one black dot. 
In the bottom panels ($t = 1.84$ Myrs), the no jet feedback (left panel) simulation has a total of 46 sink particles accounting for $\approx 1072 M_\odot$, while the jet feedback simulation (right panel) has 49 sinks for $\approx 342 M_\odot$. Both jet and no jet simulations have a total box mass of $18,000 M_\odot$.}
    \label{fig:jet_panel}
\end{figure*}

In Figure \ref{fig:jet_panel} we plot sequential projections of the entire simulation without (left column) and with (right column) jet feedback.
We show the plots at $t=0.8$ Myrs (top), which is right after the first stars form, to $t=1.33$ Myrs (middle) and $t=1.84$ Myrs (bottom). The images show up to seven levels of refinement, giving an effective resolution of $16384^3$ or a minimum cell size of $\sim 200 {\rm \, AU}$.
The black dots in the $t =  1.33 \, {\rm Myrs}$ and $ 1.84 \, {\rm Myrs}$ panels are representative of one or more sink particles. 
We state one or more, as the full box projections are zoomed out such that small clusters of two or three sinks have been plotted under one black dot. 
The black dots are also partly enlarged so that the reader may more readily compare our star formation sites between the jet and no jet simulations.
We briefly note that we do not make a specific attempt to simulate either isolated or clustered star formation.
Given our setup, we could expect to form moderate sized star clusters, of mass $\approx 2000 M_\odot$, but this does not preclude isolated star formation.

Like previous simulations, e.g. \citet{1998ApJ...504..300P,2015ApJ...800...49L}, we see that the high density regions are organized into filaments, which appear to flow into large clumps.
The clumpy regions have the highest densities and as expected form sink particles first.  The star formation efficiency (SFE) advances from $\epsilon_* \equiv M_*/M_{\rm tot} = 0$ (right top) to $0.019$ (right bottom) for the case with jet feedback, where $M_*$ is the mass in stars and $M_{\rm tot}$ is the total gas mass; for the case without jet feedback, $\epsilon_* = 0$ (left top) to $0.06$ (left bottom).
The jet feedback case has substantially reduced SFE (by a factor of 3) and the gas shows evidence of driving by jets; note the bubble near x = 1 pc, y = 5 pc in the right column at 1.84 Myrs for example, which is not present in the left column.

\begin{figure}
	\includegraphics[width=\columnwidth]{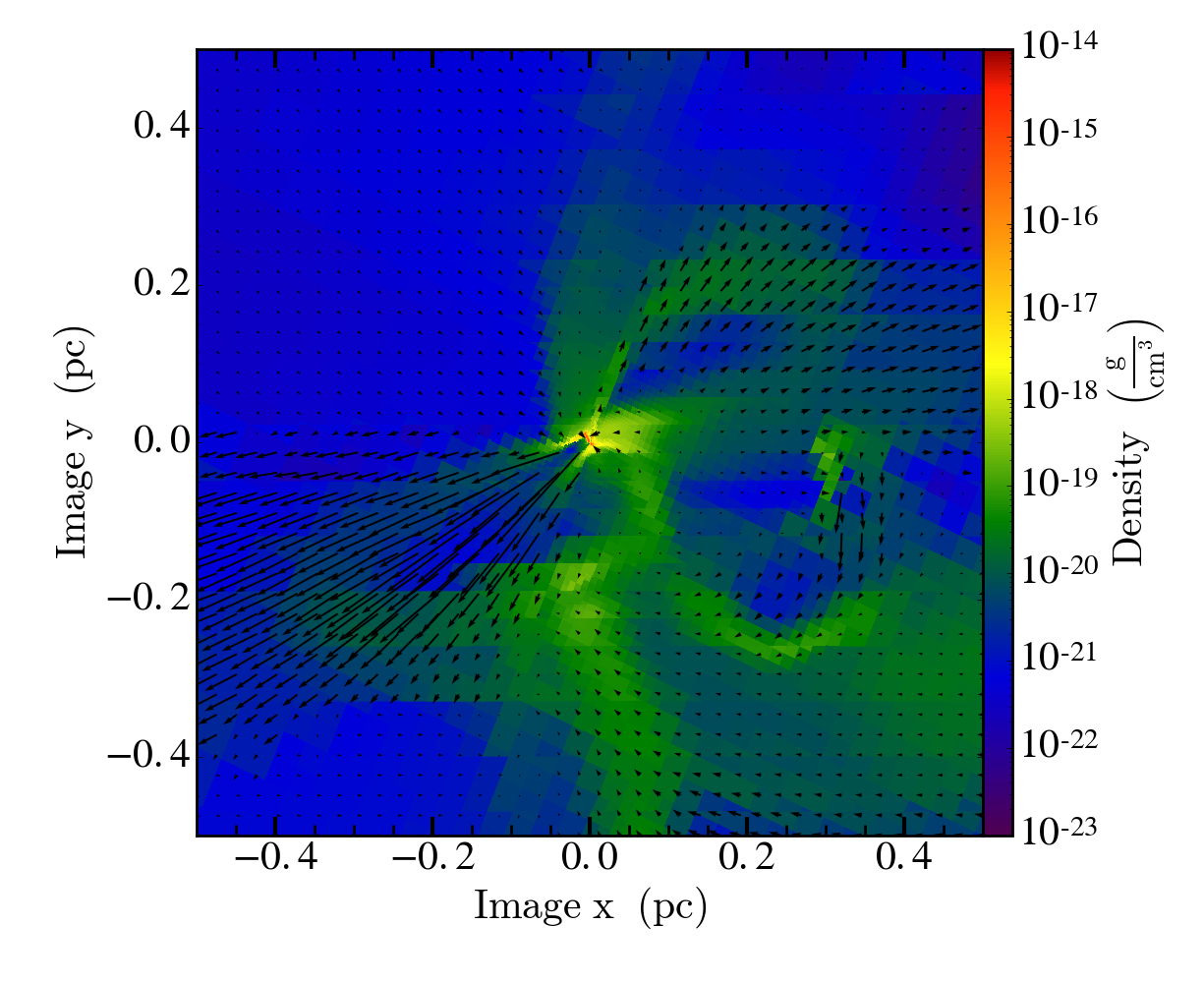}
    \caption{A slice of the density structure centreed on a sink particle of $\approx 3 M_\odot$ viewed along the angular momentum axis. Annotated in black arrows is the gas velocity, where length corresponds to relative magnitude. The jet is blowing two bubbles opening in opposite directions, partially disrupting the filament in which the star particle formed. The maximum extent of the jet is $\approx 0.5 \, {\rm pc}$ about $300,000 \, {\rm yrs}$ after formation.}
    \label{fig:jet_close_up}
\end{figure}

Figure \ref{fig:jet_close_up} displays a thin slice $1\,{\rm pc}$ on a side centred on a star particle, showing the bubble inflated by the jet from the central star.
The total extent of the bubble, which has a bi-conical shape, is roughly a parsec.
%

Despite the fact that the morphology of the filament and the surrounding gas on parsec scales is dramatically affected by the protostellar jets, we show in the next subsection that the total stellar mass accretion rate in the box is not strongly affected by the large scale effects of the jets (although it is affected, at the factor of two or three level, by the direct removal of mass from the protostar/protostellar disk by jets).

\subsection{Star Formation Rate}\label{subsec:SFR}
We begin with a discussion of the overall star formation efficiency (SFE).
MC15 developed an analytic model of turbulent collapse, motivated by the work of \citet{2015ApJ...800...49L}, who found that the SFE was $M_*\propto t^2$.
The prediction for the SFR is:
\begin{equation}
\dot{M}_* = f 4\pi r^2 \rho(r) \sqrt{\frac{GM_*(r,t)}{r}},
\label{eq:mass_accretion}
\end{equation}
where $f = 1 - \Delta M_{\rm jet} / \Delta M = 1 - f_{\rm jet}$ is the fraction of mass that accretes onto the star particle.
We have modified the expression in MC15 by including the factor $f$ to account for the ejection of mass by the jet, the expression for $\dot{M}$ can be written as $\dot{M} = f \beta M_*^{\frac{1}{2}}$, where $\beta$ is a constant in time.
We integrate from the star particles formation time to the current time to obtain:
\begin{equation}
M_*(t) = f^2 \left( \frac{\beta}{2} \right)^2 (t-t_*)^2.
\label{eq:mass_ratio}
\end{equation}

In the analytic model the $t^2$ dependence arises from the following two results.
First, the density around a collapsing region approaches an attractor solution, $\rho(r,t) \rightarrow \rho(r)$.
Second, the velocities inside the ``sphere of influence'' $r_*$, where $r_*$ is the radius at which the gas mass enclosed is roughly equal to the stellar mass, is controlled by gravity, i.e., $v_r, v_{\perp} \propto \sqrt{GM_*/r}$.
Both the fact that the density approaches an attractor, and the growth of the infall velocity at a fixed radius with $M_*(t-t_*)$  were also later verified by \citet{2017MNRAS.465.1316M} using high resolution AMR simulations.

\begin{figure}
	\includegraphics[width=\columnwidth]{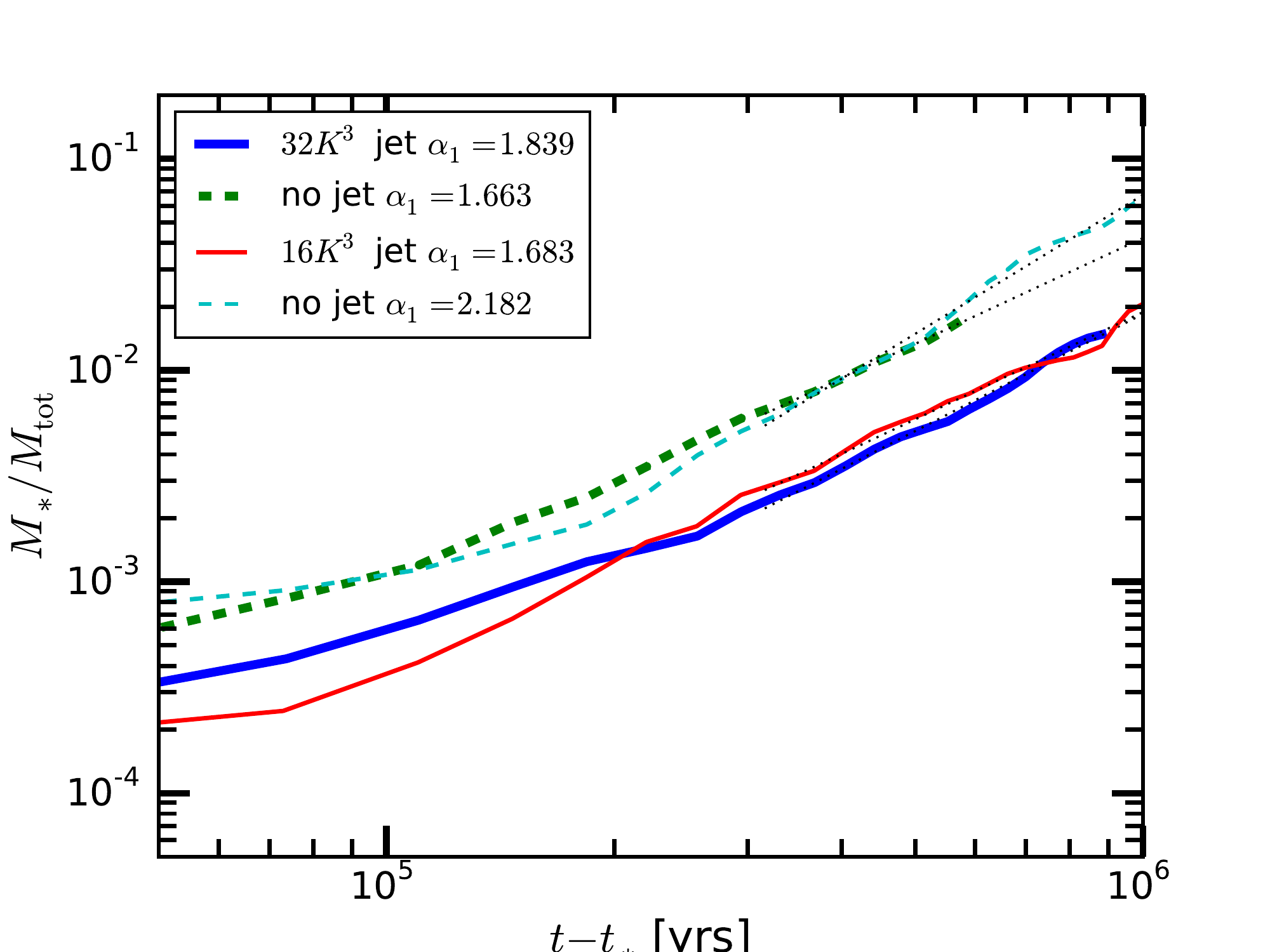}
    \caption{Star formation efficiency (SFE) as a function of time since the formation of the first star, $t-t_*$. Power law fits to the SFE for $M_*/M_{\rm tot} > 10^{-3}$ for an effective resolution of $32k^3$, $M_*/M_{\rm tot} \propto (t - t_*)^{\alpha}$, give $\alpha \approx 1.8$ and $\approx 1.7$ for the jet and no jet cases, respectively.
    We have done runs at $8k^3$, (not shown) for which the SFR does not appear to be converged.
    We note that the power law fits for both jet and no jet simulations are roughly equivalent. That is, the inclusion of protostellar jets does not affect the general dynamics of star formation. However, it is important to note that the case without protostellar jets has a larger percentage of mass in stars at any given time compared to the jet simulations. This indicates that while the inclusion of jets does not affect the power law of the accretion rate, it does reduce the mass accreted onto any given star particle at a fixed time after birth, by construction in our simulations, and by disk and/or X winds in real stars; see \S \ref{subsec:SFR}.}
    \label{fig:sfr}
\end{figure}

Does this story change in the presence of jet feedback? It appears that qualitatively it does not.  In Figure \ref{fig:sfr} we show the SFE, $M_*(t-t_*)$, as a function of time since the first star particle was formed, $t_*$.
We also report power law fits to the SFE (for $M_*/M_{\rm tot} > 10^{-3}$), for both the $16k^3$ and $32k^3$ effective resolution runs.
The fit gives $M_*/M_{\rm tot} \propto (t - t_*)^{\alpha}$ with $\alpha \approx 1.8$ for both the jet and no jet cases. If the emergence of this $t^2$ law is due to the same physics as was shown in the no-jet case, then again the density approaches an attractor solution and velocity inside of the sphere of influence of the star particle is controlled by the stellar gravity.  We demonstrate these two facts below in \S~\ref{sec:run_density} and \ref{sec:small_scale}, which show that the inclusion of protostellar jet feedback does not strongly affect the dynamics of star formation.

The normalization in Figure \ref{fig:sfr} shows that at a given time, the no jet case has approximately 2.5 times the mass of the jet case.
Calculating the ratio of the stellar masses at a fixed time after star formation for the no jet vs jet runs, we find a ratio of $2.5$ and $2.6$ for the $16 \, K^3$ and $32 \, K^3$ resolutions.
However, an examination of equation (\ref{eq:mass_ratio}) reveals that the change in the normalization is explained almost entirely by the fact that the jet case ejects $f_{jet} = 0.3$ of the mass accreted onto the star back into the interstellar medium in the form of a jet.
In particular, equation (\ref{eq:mass_ratio}) indicates that $M_* \propto f^2$, which implies that if the jet is ejecting a fraction $f_{\rm jet} = 0.3$ of the accreted mass then the star's mass is only $f^2 = (1-f_{\rm jet})^2 = (0.7)^2 \approx 0.5$ of what its mass would be if there was no jet.
MC15 and \citet{2017MNRAS.465.1316M} assumed $f_{\rm jet} = 0 \rightarrow f = 1$, so it appears that the incorporation of jet feedback into the analytic theory of MC15 can be accomplished with a simple physical parameter!

\begin{figure}
 \includegraphics[width=\columnwidth]{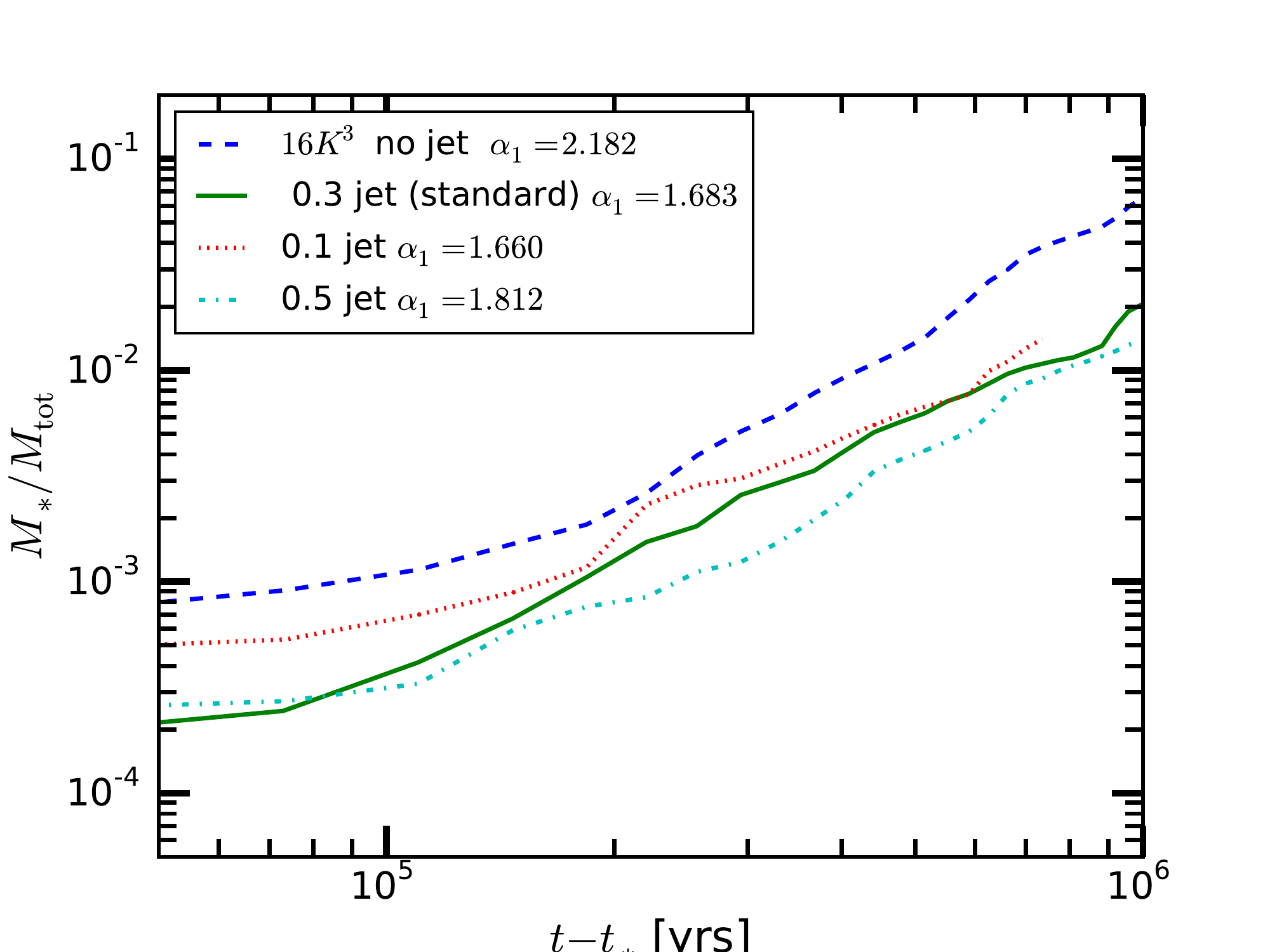}
 \caption{Star formation efficiency (SFE) as a function of time since the formation of the first star, $t-t_*$ for 4 simulations, all with an effective resolution of $16k^3$. The jet simulations differ only in the fraction of accreted mass that they eject, the total momentum is the same across all jet simulations. The blue dashed line (labeled no jet) is the same simulation as the cyan dashed in Figure (\ref{fig:sfr}) and the solid green line (labeled $0.3 {\rm \, jet}$) is the same as the thin solid red line in Figure (\ref{fig:sfr}). The dotted red line is $f_{\rm jet} = 0.1$, while the cyan dot-dashed line is $f_{\rm jet} = 0.5$. The effective resolution, jeans length and initial conditions are identical for all simulations in this plot. 
}
 \label{fig:f_jet}
\end{figure}

Figure \ref{fig:f_jet} plots the SFE for four different simulations, where we varied $f_{\rm jet}$ while keeping $p_{\rm jet}$ constant.
The ratio predicted by equation (\ref{eq:mass_ratio}) with $f = 0.7$ is $2.04$ and accounts for $75 \%$ of the difference between the two runs.
Thus the (indirect) effect of the jet on accretion onto star particle (from turbulent driving, explusion of accreting gas, etc) is $\approx 25 \%$, showing that the dynamical effect of jets on the mass accretion rate is minor.
Comparing the left to right hand columns in Figure \ref{fig:jet_panel}, this is not too surprising, since the differences are rather subtle; Figure \ref{fig:jet_close_up} shows a zoom-in on a proto-star with a jet-inflated bubble, which shows that, while the jet moves gas on scales of order a parsec, accretion continues along directions perpendicular to the jet axis.

In the next subsections we check the other predictions of MC15, $\rho(r,t) \rightarrow \rho(r)$ is an attractor solution, $u_r = \sqrt{GM_* /r}$ is set by the protostar at $r<r_*$, where $r_*$ is the sphere of influence of the star particle (MC15), and the ``raw'' mass accretion rate -- without the $f_{\rm jet}$ correction --  is the same in the cases with and without a jet.

\subsection{A fixed point attractor for $\rho(r,t)$ inside $r_*$} \label{sec:run_density}

One of the more striking findings in MC15 and confirmed in the simulations of \citet{2017MNRAS.465.1316M} was that the run of density is independent of time for $r < r_*$, where $r_*$ is the radius of the sphere which encloses a gas mass equal to $3 M_*$, i.e., the sphere of influence of the star.
Those simulations were hydrodynamic runs only, and thus the question addressed here is, do jets alter the density profile?
\begin{figure}
	\includegraphics[width=\columnwidth]{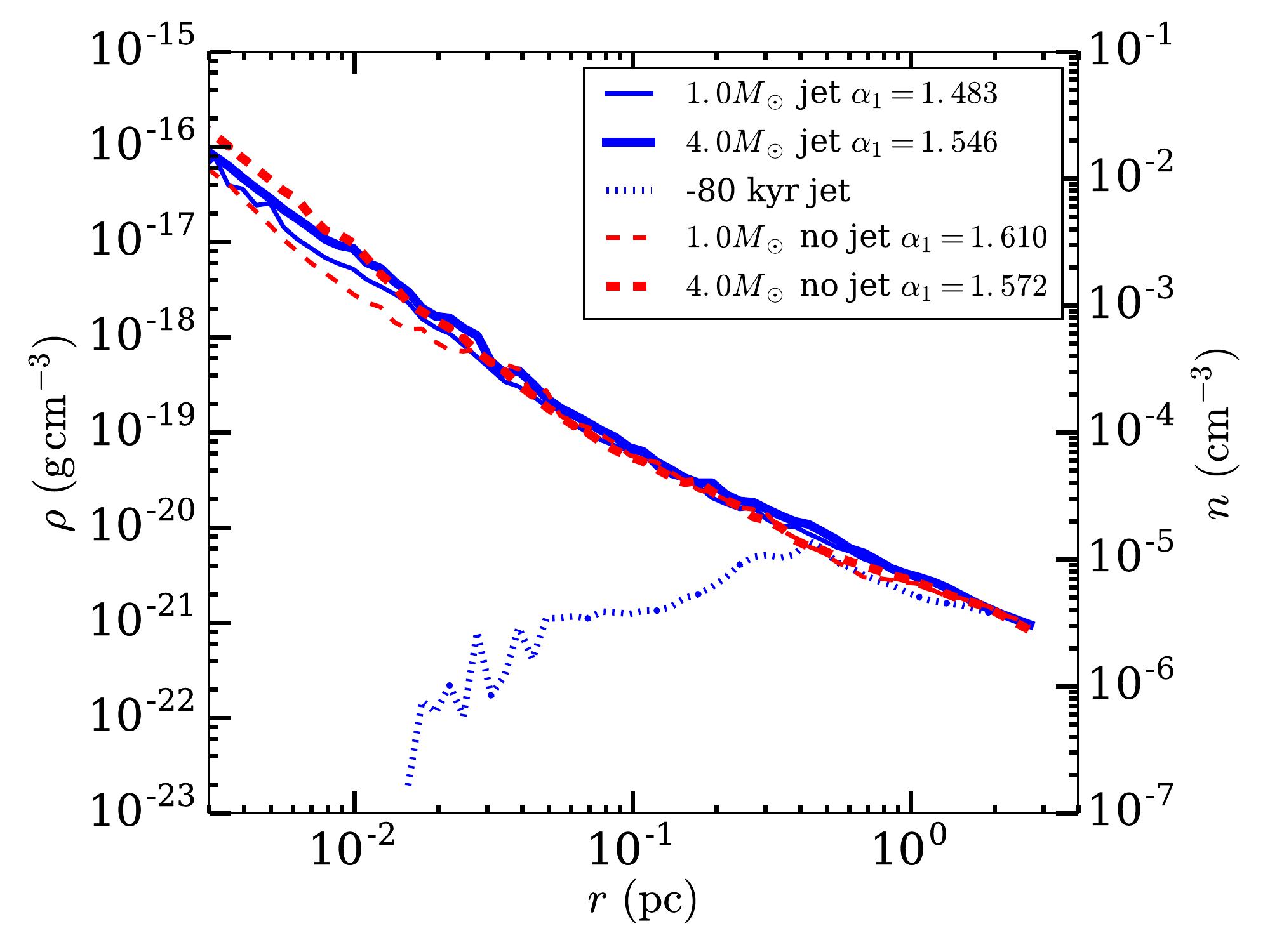}
    \caption{Density $\rho$ as a function of radius at 80,000 years prior to star formation (blue dotted line) and for $1$ (thin lines) and $4 M_\odot$ (thick lines) sink particles for both the jet (blue solid lines) and no jet (red dashed lines) simulations. In the jet case for the one solar mass stars we average over thirty six particles, and thirty particles for the four solar mass case. The corresponding no jet cases are averaged over nine and twenty three particles. Finally, the plot at 80,000 years prior to formation in the jet case is averaged over 16 particles.
    }
    \label{fig:ideal_density}
\end{figure}

In Figure \ref{fig:ideal_density} we plot the averaged number density $n$ and mass density $\rho$ as a function of $r$ for $1$ and $4 \, M_\odot$ sink particles.
The averages are over 36 and 30 particles respectively.
The plot confirms that $\rho(r,t) \rightarrow \rho(r)$ for $r_d < r < r_*$, i.e. the density is already on an attractor solution and that profile persists well after formation.
We define the accretion disk radius $r_d$ as being the radius where the circular velocity ($v_\phi$) is larger than $v_T$ and $u_r$.
The mean power-law slope of the density after the star forms is $k_\rho \sim 1.5$. In addition the density profile is the same for the jet and no-jet case confirming a major ingredient of equation (\ref{eq:mass_ratio}) showing that the results of MC15 continue to apply in the case of jet feedback.

It is important to note that the lack of change in the run of density is not due to the fact that we integrate for roughly a quarter of the global free-fall time.
We emphasize that the density can change on the local free-fall time, which is much smaller than the global free-fall time.
This can be seen from Figure \ref{fig:ideal_density}, where $\rho(r)$ changes rapidly for $r < 0.3 {\rm pc}$ before the star forms (see the dotted line).

\subsection{The infall ($u_r$), random motion ($v_{\rm T}$)  and circular ($v_{\phi}$) velocities} \label{sec:small_scale}

\begin{figure*}
    \includegraphics[width=\columnwidth]{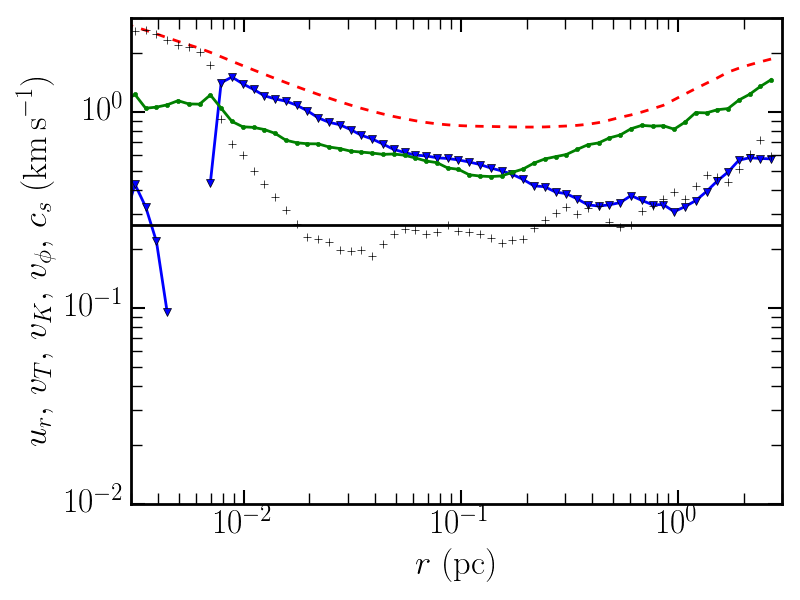}
    \includegraphics[width=\columnwidth]{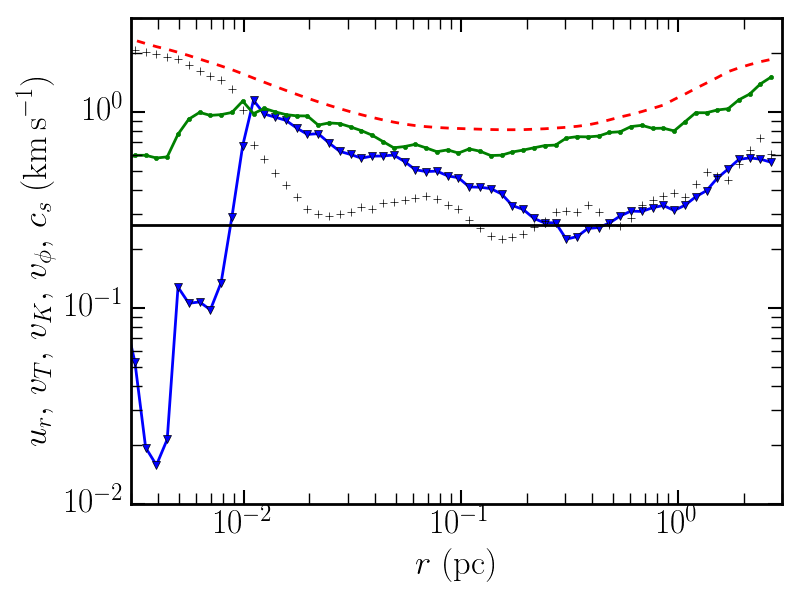}
    \caption{The left panel is the run of velocity for a $\sim 3.0 {\rm \, M_\odot}$ sink particle with jet feedback. The right panel is the particle in the same location in a simulation without jet feedback, with mass $3.8 \, M_\odot$. The sound speed is denoted by the black horizontal line while the infall velocity $ \lvert u_r \rvert$ is given by the blue triangles, connected by a solid blue line. The green circles connected by a solid green line show $v_T$ while the black crosses show the rotational velocity $v_K \equiv \sqrt{GM(< r)/r}$. Note that both $\lvert u_r \rvert$ and $v_{\rm T}$ increase with decreasing radius for $10^{-2}{\rm\, pc}<r<10^{-1}{\rm \, pc}$. }
    \label{fig:velocity_profile}
\end{figure*}

Figure \ref{fig:velocity_profile} shows the infall velocity $u_r$ (blue triangles connected by a solid blue line), the random velocity $v_{\rm T}$ (green dots connected by a green line), and the rotational velocity $v_\phi$ (black plus signs) as a function of radius centered around a specific sink particle.
The red dashed line depicts $\sqrt{GM(<r)/r}$, while the black horizontal line shows the sound speed $c_s = 0.265 {\rm \, kms^{-1}}$.
We calculate each of these velocities in the same manner as \citet{2017MNRAS.465.1316M}, with the following modification for the jet case: when calculating these otherwise spherical shell averaged quantities, we remove the cells that sit within the opening angle of the jet.
See Appendix \ref{sec:jet_cone_sub} to see how this subtraction affects the radial profiles of all the velocities.

The left panel in Figure \ref{fig:velocity_profile} shows the velocities associated with a three solar mass sink particle in a simulation that included jet feedback.
The right panel is of a sink particle in a simulation without jet feedback that formed within a hundredth of a parsec of the same location within a few thousand years of the same time as the particle in the jet case.
That particle is roughly $3.8 M_\odot$.
While we can not make a direct comparison of the magnitude of the velocities, due to the different masses, we can compare the trends in the velocities to the $\sqrt{GM(<r)/r}$ velocity in each panel. We see similar behavior in both jet and no-jet simulations, with some expected differences.

First, what one immediately notices in each case (with and without jets), is that both $\lvert u_r \rvert$ and $v_{\rm T}$ decrease with decreasing radius down to $r \approx 0.1 {\rm pc}$ then increase with decreasing radius, until running into the accretion disk around the sink particle, at $r_d \approx 5\times 10^{-3} {\rm pc}$.
This indicates that jets by themselves do not change the general dynamics outside and inside the stellar sphere of influence.
 Both $\lvert u_r \rvert$ and $v_{\rm T}$ increase inward of $r_*$. This was seen by \citet{2017MNRAS.465.1316M} for the no jet case.
It is also consistent with the rapid increase in turbulent energy density with decreasing radius seen in \citet{2017ApJ...838...40M}, although their simulations halted when the first star particle formed (and hence did not include any protostellar jet feedback).

However, the inclusion of jets does affect the ratio of the random motion velocity to the fiducial velocity $\sqrt{GM(<r)/r}$.
In the case with jets, $v_{\rm T}$ remains much closer to $\sqrt{GM(<r)/r}$ for all radii larger than the disk radius as compared to that ratio for the no-jet simulation.
This is not an unexpected result, given that jets are injecting momentum back into the gas surrounding the collapse; this momentum deposition pumps up the random motions of the gas and thus slows down the in-falling gas.
Even though the jets are boosting the random motion velocity, inside the stellar sphere of influence, $\lvert u_r \rvert$ still increases with decreasing radius. That is, the gravity of the star dominates the dynamics.


Finally, we note that the infall velocity is $\sim 25 - 30 \%$ of the free-fall velocity over all radii less than a parsec. This observation shows that this system is not in hydrostatic equilibrium. The run of density versus radius is similar to that in Figure \ref{fig:ideal_density}.

\subsection{Average mass accretion rate $\langle \dot{M}(r, t) \rangle_{\rm stars}$ for jet and no jet sink particles} \label{subsec:mdotstars}

\begin{figure}
	\includegraphics[width=\columnwidth]{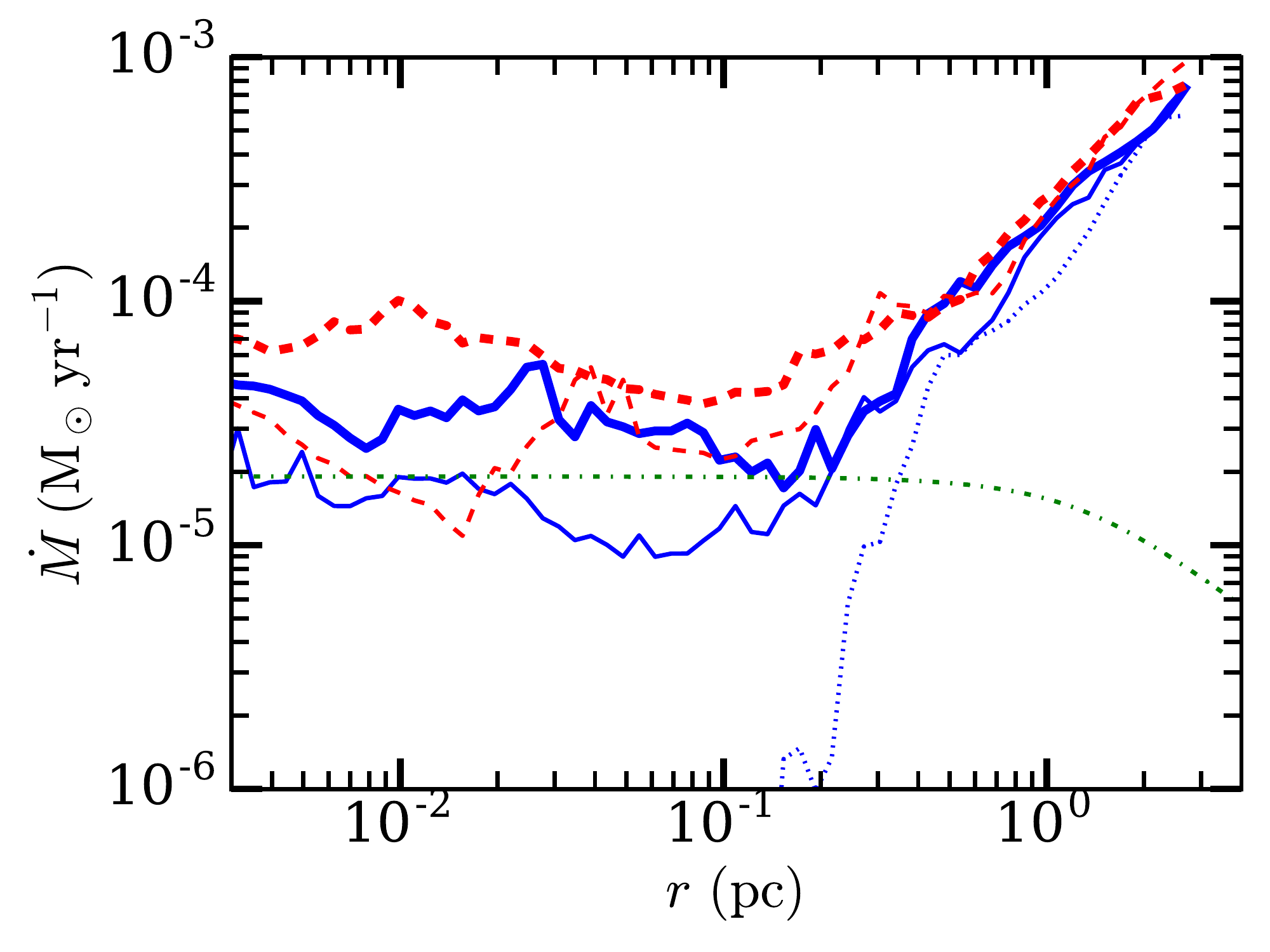}
    \caption{The average run of $\dot{M}$ for $1$ and $4 \, M_\odot$ sink particles in simulations with and without jets. The blue solid lines are the jet simulations, while the red dashed lines are for the no jet case. The thin lines correspond to the $1 \, M_\odot$ averages, while the thick lines represent the $4 \, M_\odot$ averages. The thin blue dotted line is the average 80,000 years prior to formation in the jet case. The average, for the jet case, is over 16, 30 and 36 sink particles for the prior to formation, $1$ and $4 \, M_\odot$ respectively. For the no jet case, the average is over 9 and 23 sink particles for the $1$ and $4 \, M_\odot$ respectively. In all cases, the accretion rate is about an order of magnitude  lower at small radii (say $10^{-2} \rm{pc}$) than at $1 \rm{pc}$. At both masses, and prior to formation, the accretion rate at $1 \rm{pc}$ exceeds that at all smaller radii, showing that the collapse is outside-in, not inside-out. As an example of an inside-out collapse, we show the accretion rate for the \citet{1977ApJ...214..488S} model (the green dot-dashed line) for a star of a solar mass with Shu's parameter $A = 3.501$.  
    }
    \label{fig:mdot_shu}
\end{figure}

The final prediction that we can check is that $\dot{M}(r,t)$ is independent of $r$ for $r < r_*$ and that the jet and no-jet cases have (roughly) the same $\dot{M}$.
In Figure \ref{fig:mdot_shu}, we show the average mass accretion rate $\dot{M}$ as a function of $r$ for $1$ and $4 \, M_\odot$ sink particles comparing between the jet and no jet simulations.

We note that while the $\dot{M}$ profile is flat at small radii it does increase over time: the profile at $4 \, M_\odot$ inside of $r_*$ is larger than that of the $1 \, M_\odot$ profile for both jet and no jet simulations.
We contrast this with an inside-out collapse model, which we exemplify using a \citet{1977ApJ...214..488S} solution, green dashed line obtained by directly integrating equations (11) and (12) of \citet{1977ApJ...214..488S} at a fixed time.

We also note that both the jet and no-jet case settle onto an average $\dot{M}$ that is roughly the same for the  $1$ and $4 \, M_\odot$ sink particle case.  Hence the ``raw'' rate is the same in the jet and no-jet cases as implicitly required by equation (\ref{eq:mass_ratio}).
%


\subsection{Jet Momentum Deposition} \label{subsec:jet_mv}
\begin{figure}
	\includegraphics[width=\columnwidth]{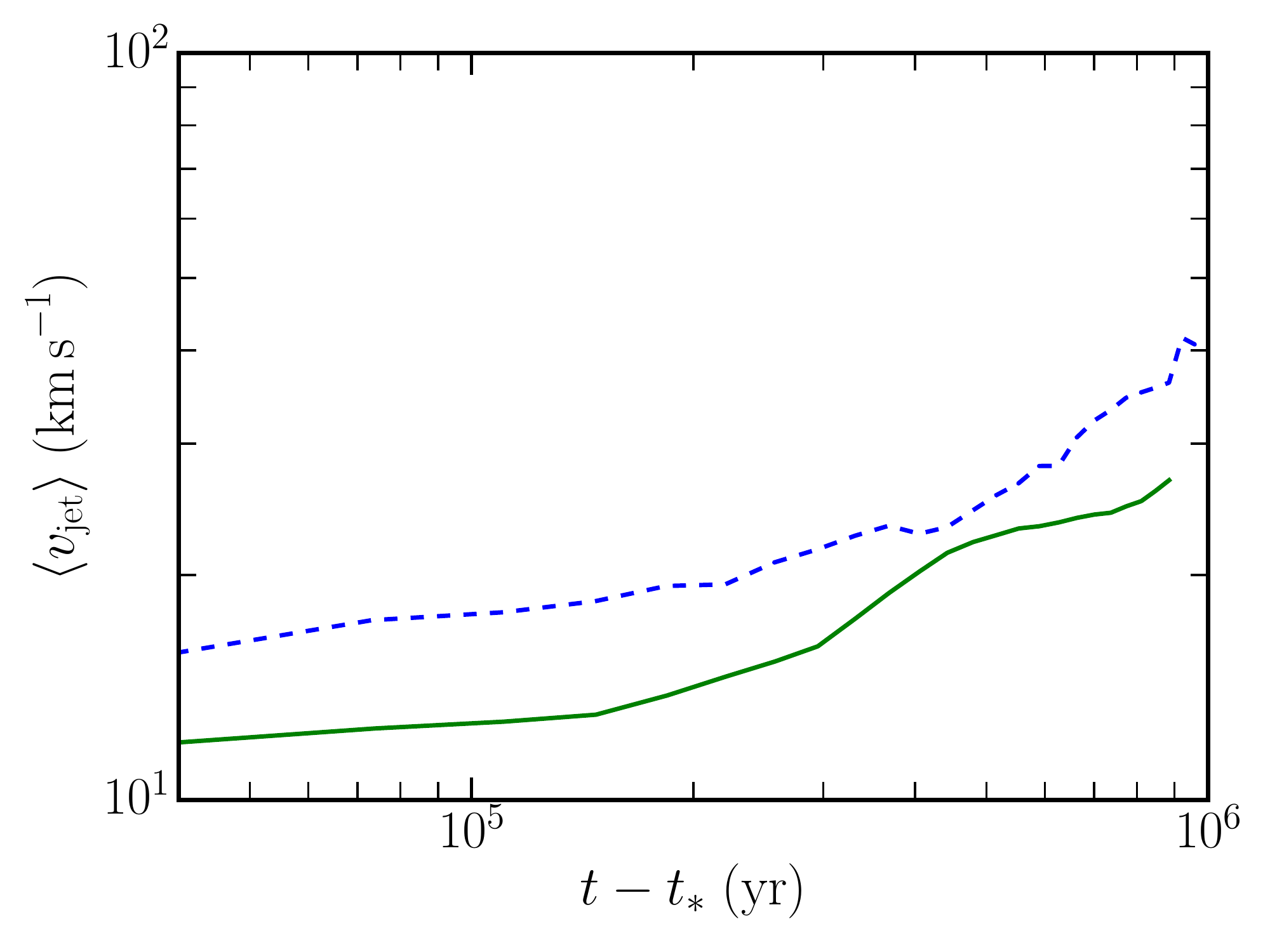}
    \caption{The average jet momentum per stellar mass as defined in equation (\ref{eq:avg_jet_momentum}) as a function of time since the first sink particle forms ($t-t_*$). The solid green line depicts the $32 \, {\rm K^3}$ run while the dashed blue line depicts the $16 \, {\rm K^3}$ result. Both show an increase in $\langle v_{\rm jet} \rangle$ with increasing time. This occurs because as time increases, the star particles gain mass, and thus have higher escape velocities and launch faster jets. The $16 \, {\rm K^3}$ run produces higher jet momenta per stellar mass because it forms fewer stars that are more massive, due to its lower resolution. Note however that the total stellar mass in the two runs is converged (see Figure \ref{fig:sfr}).}
    \label{fig:avg_jet_velocity}
\end{figure}

In Figure \ref{fig:avg_jet_velocity} we plot the mass weighted average of the jet velocity for all sink particle jets over time.
It has been calculated following the definition in \citet{2000ApJ...545..364M}
\begin{equation}
\langle v_{\rm jet} \rangle \equiv p_{\rm jet} / M_* ,
\label{eq:avg_jet_momentum}
\end{equation}
where $M_*$ is the total stellar mass and $p_{\rm jet}$ is the total momentum ejected by all the jets up to the current time.
\citet{2000ApJ...545..364M} estimate (observationally) that $\langle v_{\rm jet}\rangle \approx 40 \, {\rm km \, s^{-1}}$.
The figure shows that the average velocity is increasing with time because the stars are accreting more mass, and their radii are contracting.
By the end of our run, the average velocity is in the range of that estimated by \citet{2000ApJ...545..364M}.

\begin{figure*}
	\includegraphics[width=0.49\textwidth]{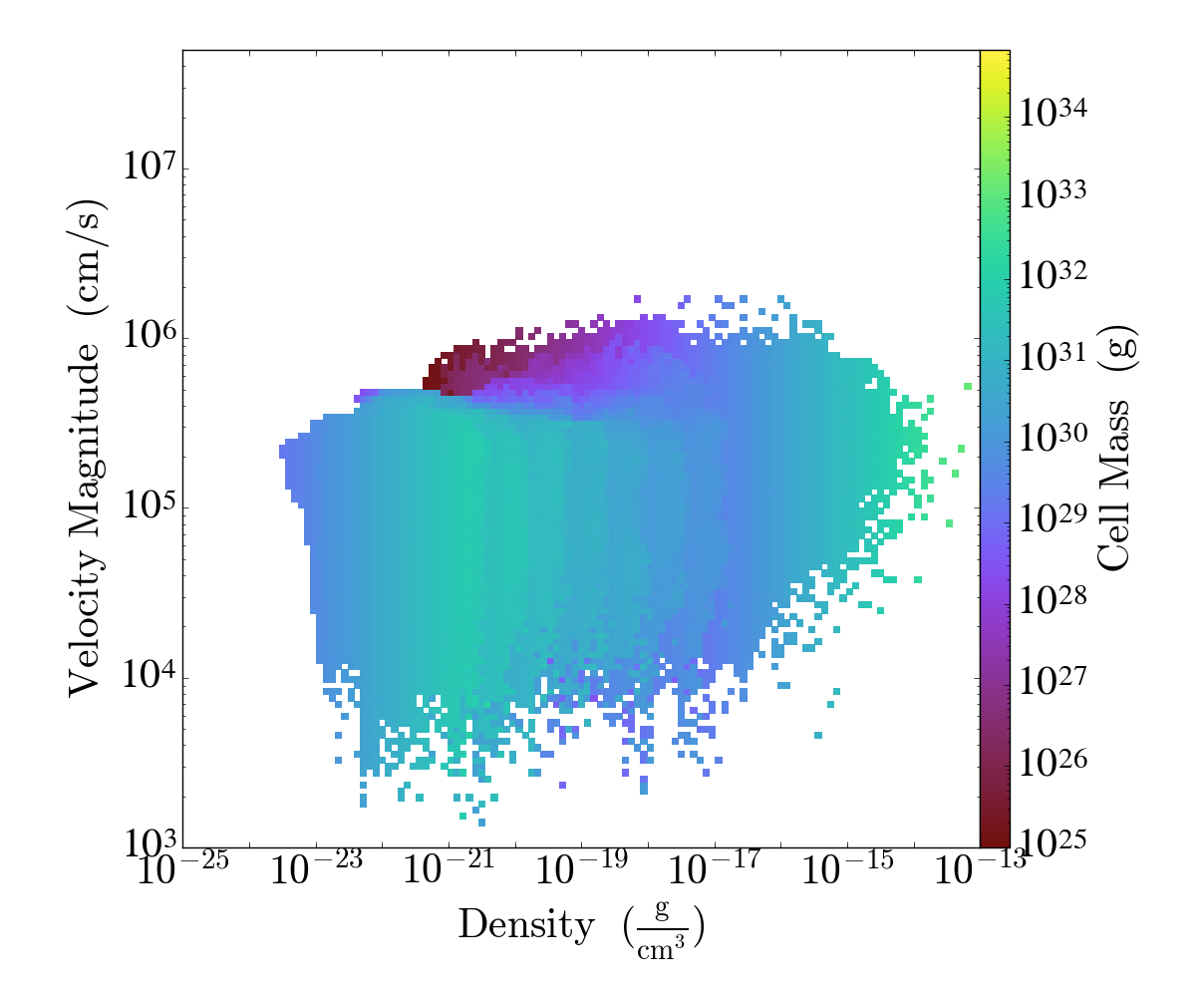}
	\includegraphics[width=0.49\textwidth]{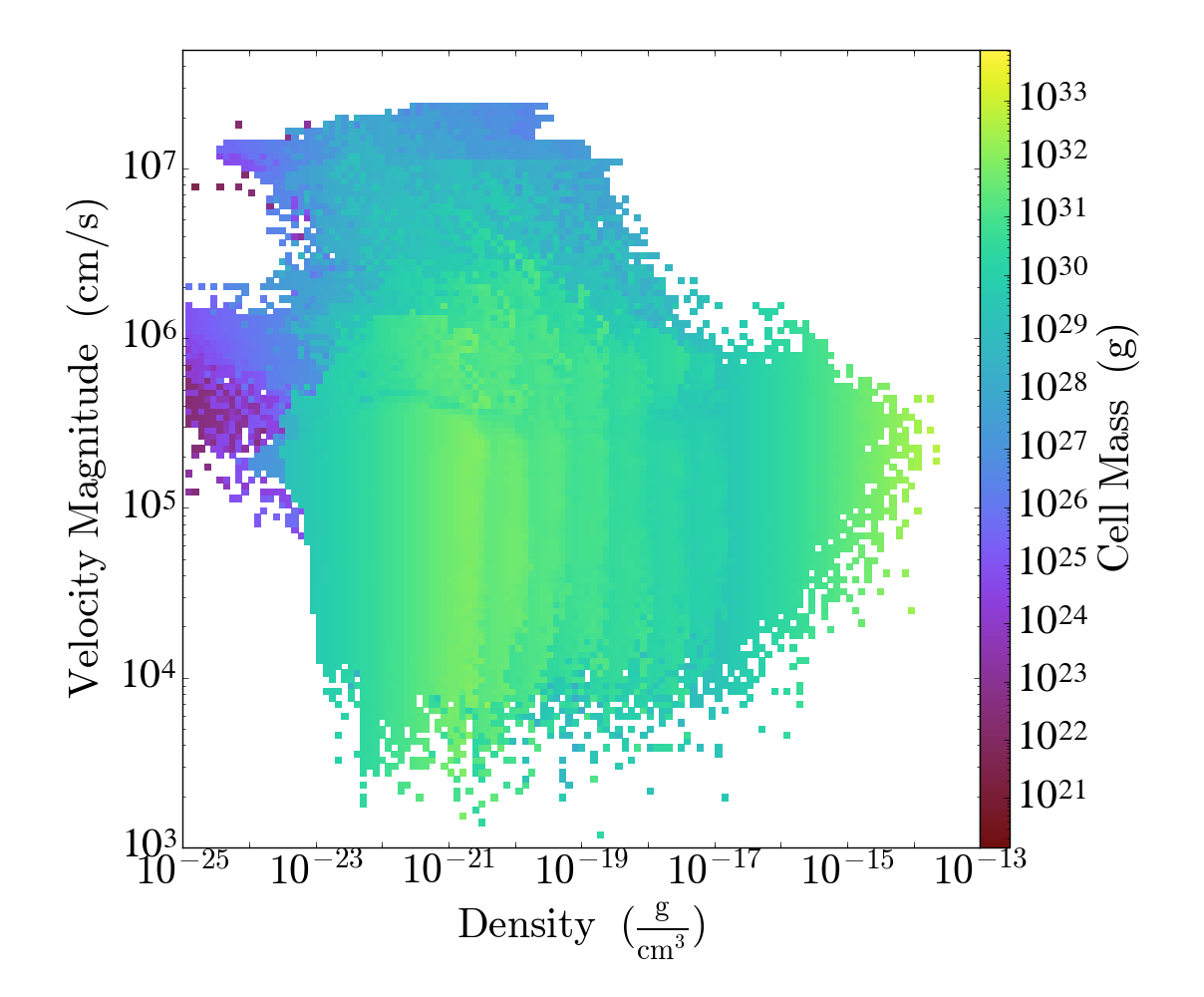}
    \caption{ Phase plot of density vs velocity for the no-jet case (left) and the jet case (right) at $t-t_*=0.48$ Myrs. The colourmap denotes the total mass at each point in density-velocity space.  The distribution at the high density end is set by gravity both accelerating the infall velocity and adiabatically heating the turbulent velocity, and is similar for both the no-jet and jet cases.  There is, however, a clear difference between the no-jet and jet cases, namely, the substantial amount of gas above $10\,{\rm km\,s}^{-1}$ in the jet case. While there is a large distribution in velocity, this has only a $25\%$ effect on the mass accretion rate (see Figure \ref{fig:sfr} and \S~\ref{subsec:SFR}).}
    \label{fig:v_phase_plot}
\end{figure*}

We note that despite the relatively low value of the mass weighted average jet velocity over the entire box, the actual jet velocity is not necessarily slow.
In Figure \ref{fig:v_phase_plot}, we show the phase plot of density vs velocity for the no-jet case (left) and the jet case (right) at $t-t_*=0.48$ Myrs, which corresponds to the middle panels of Figure \ref{fig:jet_panel}.
The colourmap denotes the total mass of each point in density-velocity space.
At the high density end, note that the velocities are substantially larger than the sound speed.
This is due to infall stirring up $v_{\rm T}$ and $\lvert u_r \rvert$ as we saw in the velocity profiles above.
A comparison between the no-jet and jet case on the high density end looks unsurprisingly similar.
The jets are directed into lower density regions, not into the accretion disk and thus do not affect the high density regions.

However, we note that the major difference between the no-jet and jet cases is the substantial amount of gas above $10\,{\rm km\,s}^{-1}$ and that the density of this gas is around 10x the mean density $3\times 10^{-22}\,{\rm g\,cm^{-2}}$.
In fact, the jet case has material moving above $100\,{\rm km\,s}^{-1}$.

\subsection{Jets mainly drive small scale turbulence} \label{sec:jets_drive_small}
\begin{figure}
	\includegraphics[width=\columnwidth]{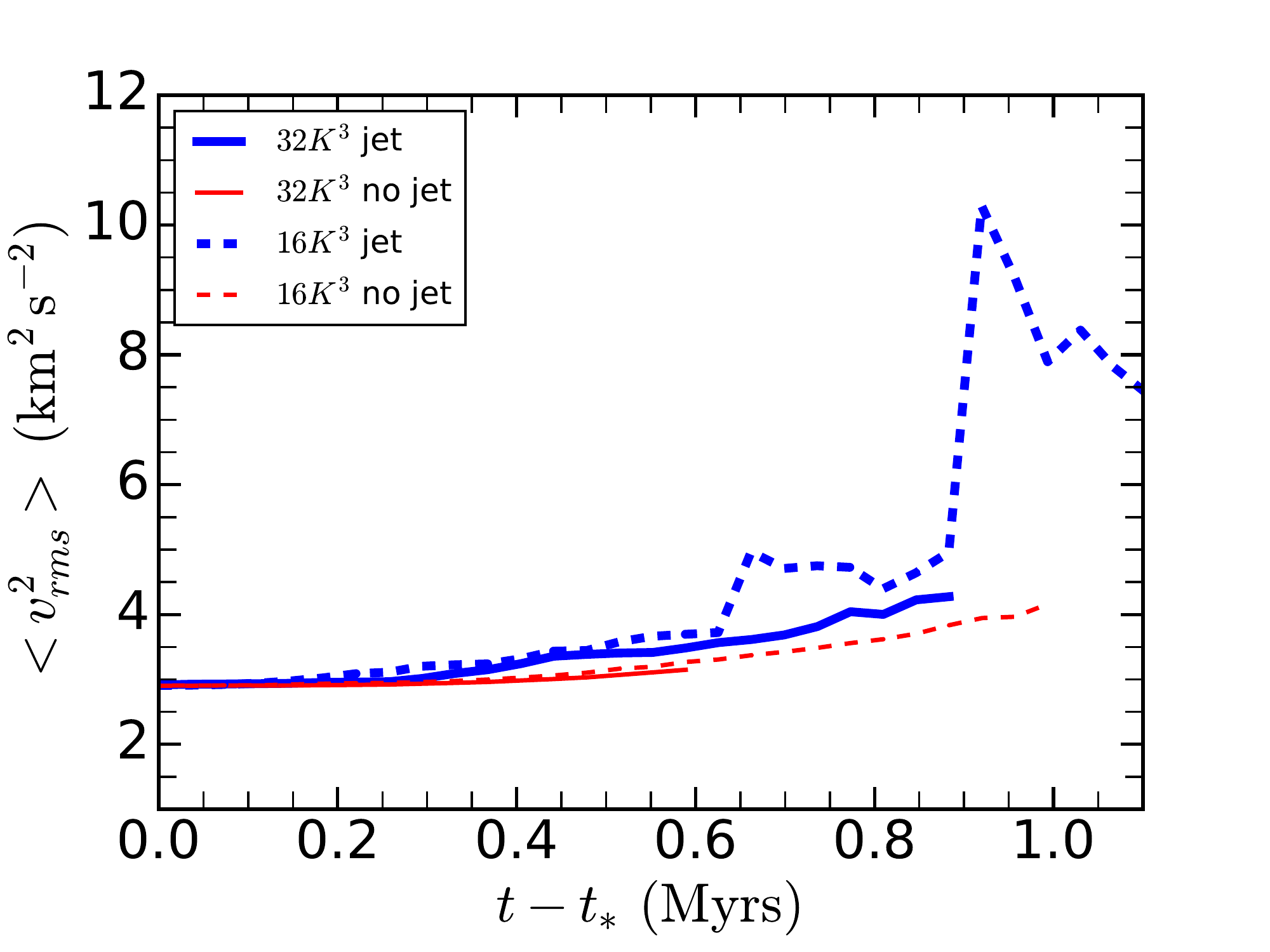}
    \caption{The mass averaged velocity dispersion squared in the simulation volume plotted as a function of time since gravity was turned on.
The thick and thin lines show the jet and no jet case respectively. The solid lines show the $32 k^3$ result and the dashed lines the $16k^3$ result.
The increase seen in the no jet case starting at $ t \approx 1 \, {\rm Myrs}$ results from the gravitational collapse driving random motions.
The excess seen over this background in the thick lines is the result of the jets driving random motions.
Both the sharp increase and the sharp decrease in the latter case alert us to the fact that this driving occurs on relatively small scales (of order $1\, {\rm pc} $): see Figure \ref{fig:jet_close_up}.
}
    \label{fig:jet_driven_turb}
\end{figure}

Figure \ref{fig:jet_driven_turb} displays the mass averaged velocity dispersion in the simulation volume plotted as a function of time since gravity was turned on.
The solid lines are the $32\, k^3$ simulation, while the dashed lines show the $16\, k^3$ simulation data.
The thick and thin lines designate the jet vs no jet cases respectively.
Looking at the no jet cases (at both resolutions), one sees an increase in the total velocity dispersion in the box from $3 {\rm km^2 \, s^{-2}}$ to $4 {\rm km^2 \, s^{-2}}$ by $t-t_* \approx 0.8 {\rm Myrs}$.
This increase is being driven by the gravitational collapse as the first stars begin to form.
The jet cases' increase over this gravitational driving indicates that the jets do stir up the surrounding medium.
It is important to note however, that this jet driving occurs on relatively small scales (of order $\sim 1 \, {\rm pc}$).
This can be inferred by the rapid increase and decrease in the plotted velocity dispersions for the jet cases.
In Figure \ref{fig:jet_close_up} we presented a postage stamp shot of a star particle with $\approx 3 \, M_\odot$ that has cleared out a region on either side of the star particle nearly half a parsec in length.
%

\begin{figure}
	\includegraphics[width=\columnwidth]{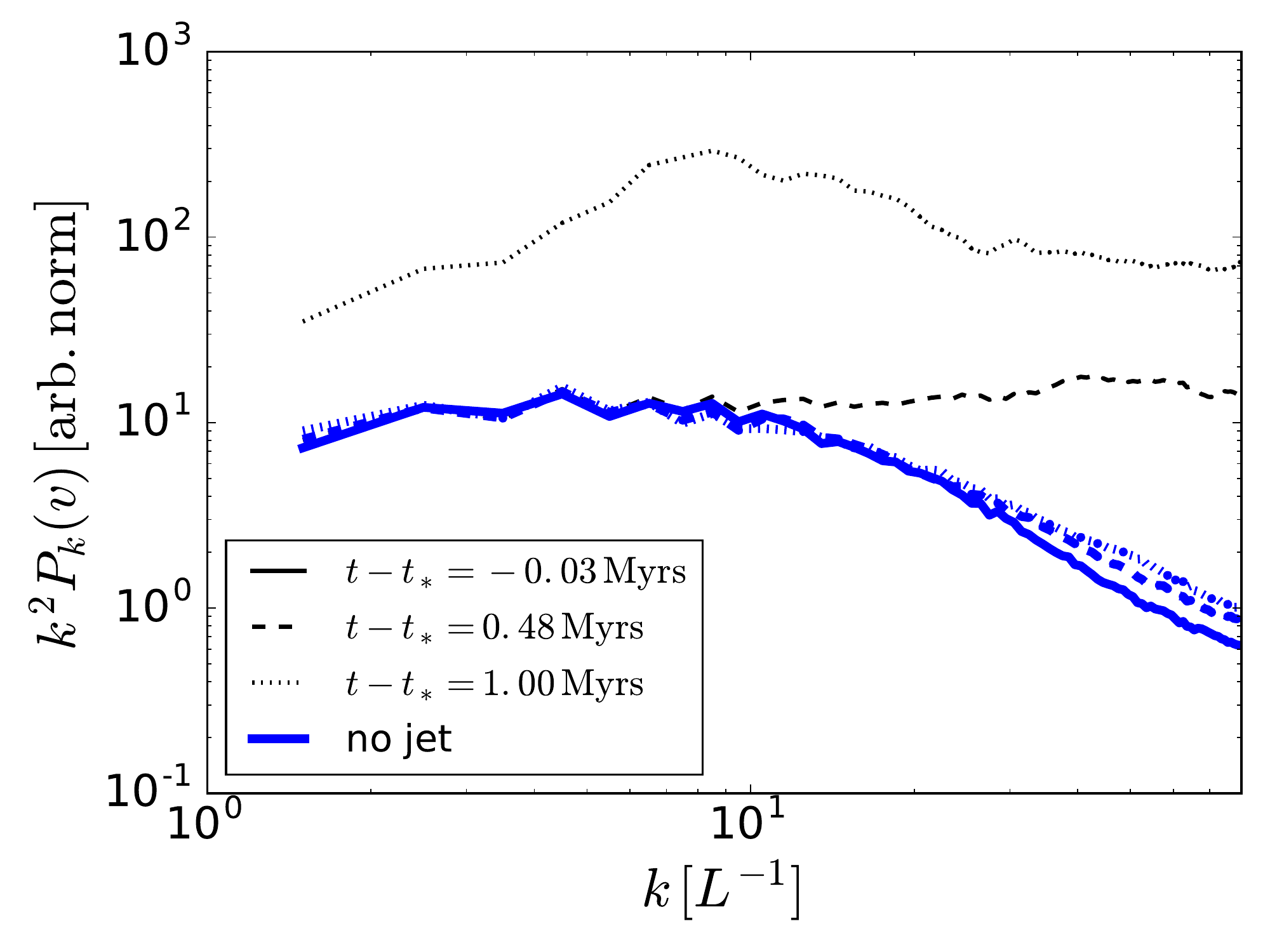}
    \caption{Compensated velocity power spectrum, $k^2P_k(v)$ as a function of wave vector, $k$ at $t-t_* = -0.03$ (solid lines), $0.48$ (dashed lines), and $1$ Myrs (dotted lines) for the jet (thin lines) and no-jet (thick lines) cases.  To perform the fast-Fourier transform, we map the simulation volume to a $256^3$ grid.  The $t-t_* = -0.03$ Myrs lines for the jet and no-jet case is the same as the spectrum is derived from stirred turbulence.  The no-jet power spectrum does not substantially differ in time as stars form, but the jet case does. In particular the flatness of the power spectrum at large $k$ at $t-t_*=0.48$ and $1$ Myr is due to the presence of velocity delta-functions (at the resolution of $256^3$) in the simulation volume.  Additionally, the jet drives an uptick in power at $kL\sim 8$, a spatial scale of 2 pc, at $t-t_*=1$ Myr.
}
    \label{fig:power_spectrum}
\end{figure}

The effect of jet driving in the velocity dispersion in the box can also be seen from looking at the power spectrum of the velocity.
In Figure \ref{fig:power_spectrum}, we plot the velocity power spectrum for $t-t_* = -0.03$ (solid lines), $0.48$ (dashed lines), and $1$ Myrs (dotted lines) for the jet (thin black lines) and no-jet (thick blue lines) cases after mapping the simulation volumes to a $256^3$ grid (to perform fast-Fourier transforms).
These power spectrum correspond exactly to the top, middle, and bottom panels of Figure \ref{fig:jet_panel}.
As expected the $t-t_* = -0.03$ Myrs lines for the jet and no-jet case lines up exactly because they start from the same initial conditions.
Moreover, the spectrum follows $P_k\propto k^{-2}$ as expected for Burgers turbulence up to where it begins to be cutoff around $k\approx 10 L^{-1}$.
For the no-jet case (thick blue lines), the power spectrum does not substantially differ in time as stars form.
This is due to the fact that the timescale that we are looking at is smaller than the crossing time of the box.
However, the jet case shows substantial deviation.
At $t-t_* = 0.48$ Myrs for the jet case (thin black dashed line), we note that the power spectrum is remarkably similar to $P_k \propto k^{-2}$ even beyond the cutoff.
In this case, this is not due to Burgers turbulence, but rather to the presence of delta-functions (at the resolution of $256^3$) in velocity in the simulation volume.
This is undoubtably due to the narrow (on the scale of $256^3$) protostellar jets. We caution the reader against the implication that this small scale velocity structure translates to turbulence in the star forming clumps. Figure \ref{fig:power_spectrum} implies that the velocity structure is small scale, but does not imply that this structure is associated with the star forming regions.  In particular, jets might induce turbulence at their working surfaces which would be at a significant distance from their launching sites.

At $t-t_* = 1$ Myrs, the jet case (thin black dotted line) show even greater deviation.  Here the velocity power spectrum is larger than it was initially.  On the larger scales, it has increase by a factor of approximately 3 and again remains remarkable flat for the same reasons discussed above.  However, at $kL\sim 8$, we note a bump in the power spectrum, which contains the bulk of the energy.  The scale of this bump is at $\sim L/8 = 2$ pc and so shows that the effect of jets is mainly on the few parsec scales in our simulation.  Moreover, the decline toward larger scales from this scale may be an indication of an inverse cascade.

\section{Discussion}\label{sec:discussion}


Given our initial set up of a box with sides $16$ pc in length, with a mach number of $\approx 9$ and a virial parameter of order unity, we could expect to form moderate sized star clusters. However, this does not preclude isolated star formation.
What we see is $\approx 180 M_\odot $ in stars, in a filament/clump $\approx 4$ pc in length and a clump gas mass of $\approx 3500 M_\odot $. The overdense regions that we see correspond to molecular clumps with a few thousand solar masses formed inside of our parent molecular cloud.

We have shown that protostellar jets do not strongly affect the dynamics of accreting gas.
For example, the infall and rotational velocities both show similar behavior between the no jet and jet runs; the random velocity is also similar, once the region containing the jet cone has been excised (see Figure \ref{fig:ideal_velocity} and appendix \ref{sec:jet_cone_sub}).
Similarly, the run of density approaches an attractor solution regardless of whether or not there are jets associated with the sink particle.

The fact that $|u_r|$ increases inward for $r<r_*$, while increasing outward for $r>r_*$, combined with the time-independence of the density profile (once a density peak forms) implies that the mass accretion rate $\dot M(r,t)$ for individual stars is flat for $r<r_*$ and increasing with radius for $r>r_*$. This is demonstrated in Figure \ref{fig:mdot_shu}, which also shows the contrasting behavior for an inside-out collapse model, that of \cite{1977ApJ...214..488S}.
This is the signature of outside-in collapse, which occurs in runs both with and with out jet feedback.

We have seen that jets do create cavities in the surrounding gas, changing the morphology of the gas on parsec scales.
However, as we discussed in \S \ref{sec:jets_drive_small}, while the jets and the expanding bubbles they inflate do drive turbulence, the outflows and bubbles do not have a large effect on the mass accretion rate.
The jets do reduce the mass accretion rate, by a factor of about 2.5 in our simulations, but the bulk of this reduction is simply due to the jet ejecting mass from the star particle.
In our simple jet model, this includes mass that in reality would be ejected from the associated protostellar disk.

We showed that the jet driven turbulence is on small scales (1 pc) compared to the scale of our simulation box (16 pc) or compared to GMC sizes (tens of parsecs).
This ensures that the jet driven turbulence from an individual star decays more rapidly (see Figure \ref{fig:jet_driven_turb} and \ref{fig:power_spectrum}) than the eddy turnover time of the simulation box, or of the host GMC in a real galaxy.

\begin{figure*}
	\includegraphics[width=0.49\textwidth]{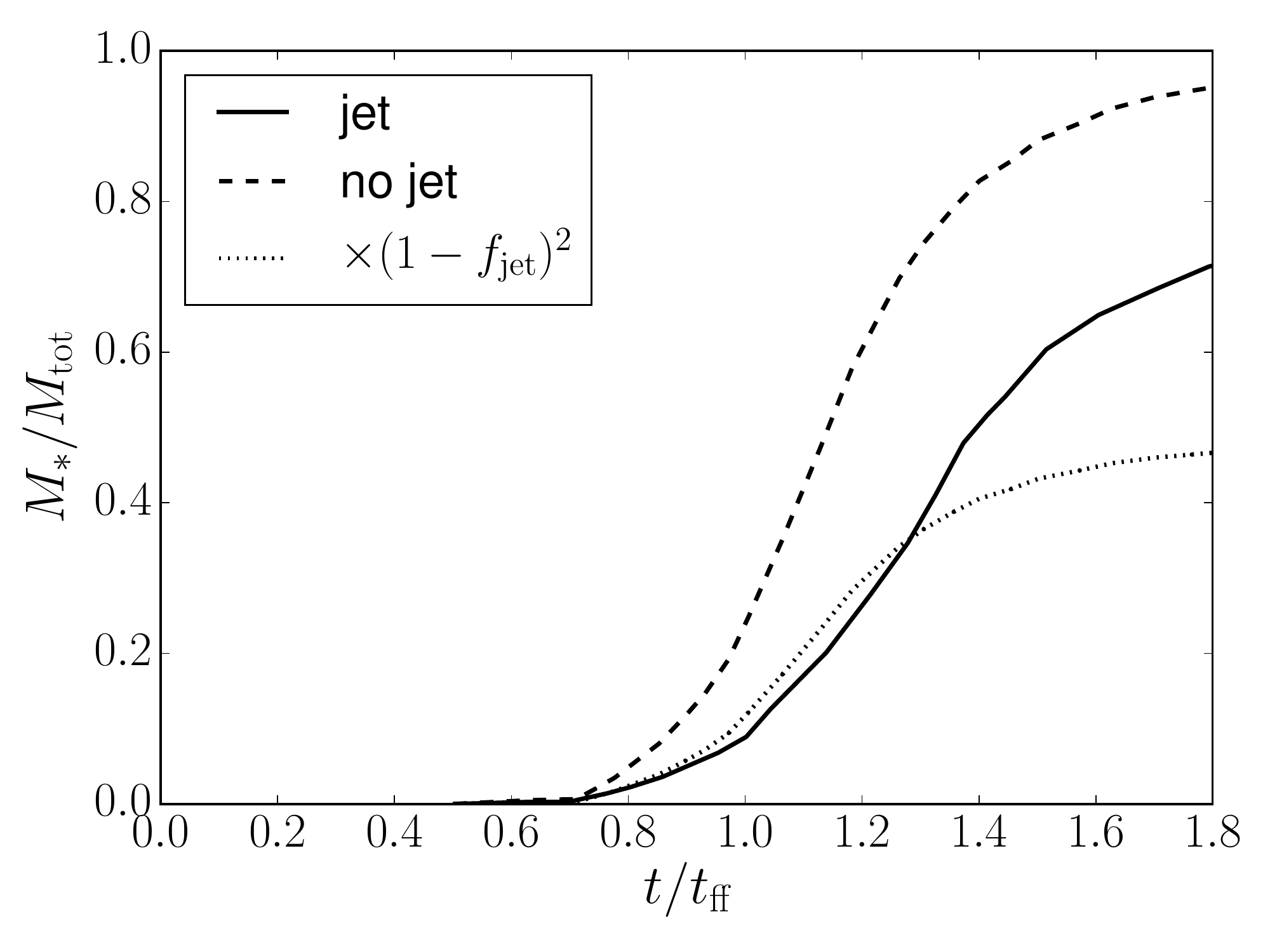}
  \includegraphics[width=0.49\textwidth]{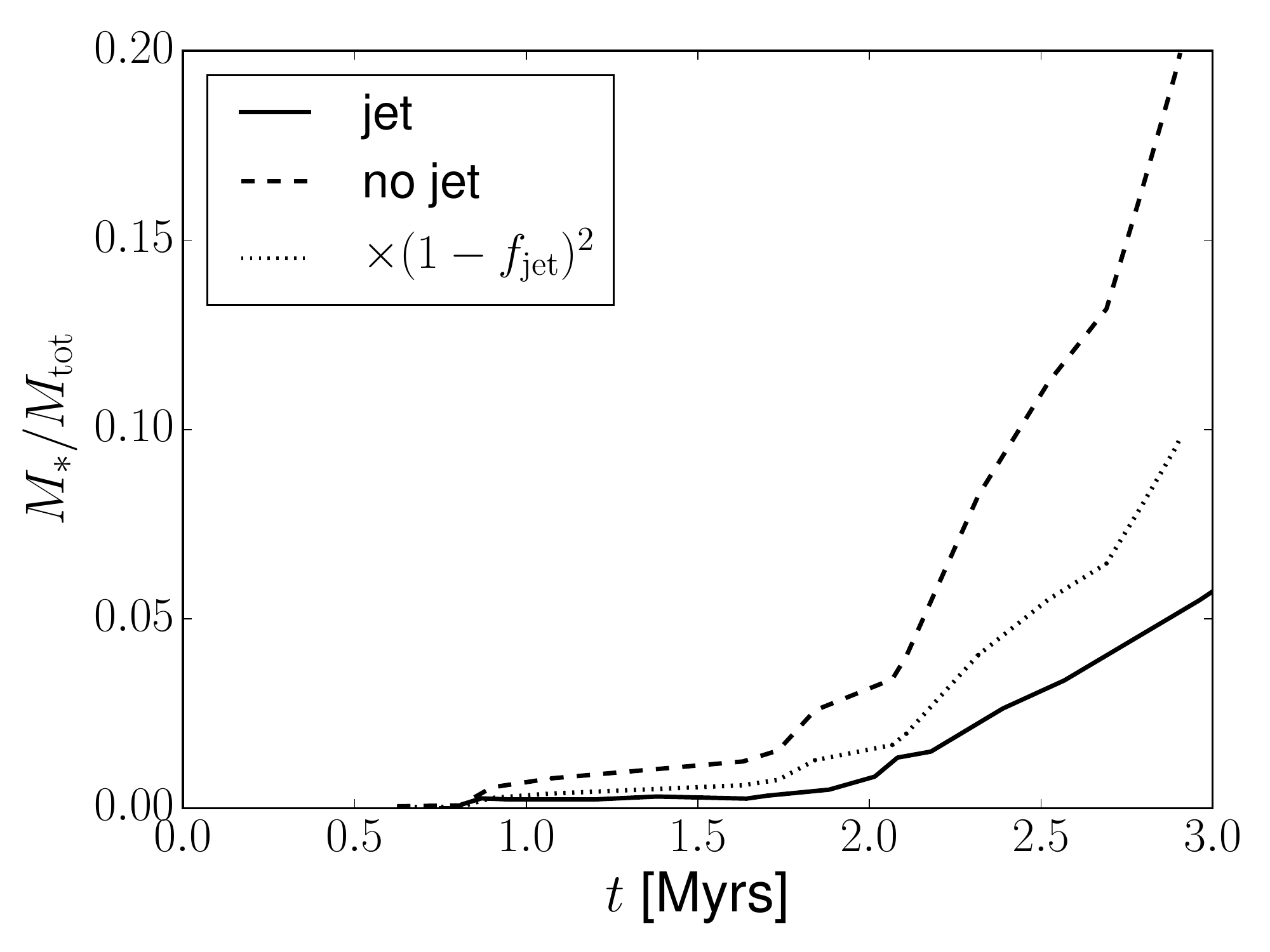}
    \caption{Left plot: the SFE, $M_*/M_{\rm tot}$, as a function of time (measure in free-fall times) for the jet (solid line) and no-jet (dashed line) cases from \citet{2014ApJ...790..128F} Figure 9  (N.B. HD simulation). 
Right plot: the SFE as a function of time (measure in Myrs) for the jet (solid line) and no-jet (dashed line) cases from \citet{2015MNRAS.450.4035F} Figure 2 (N.B. MHD simulation). 
In both plots we have rescaled the no-jet case by $f=(1-f_{\rm jet})^2$, which is set to $f_{\rm jet} = 0.3$. 
In the left plot the rescaled no-jet case follows the jet case fairly well up to about an SFE of about 0.4 . The rescaled no-jet case at this point rescales the no-jet case for $M_*/M_{\rm tot}\approx 0.8$, where the accretion onto star particles is starved due to depletion of gas. 
In the right plot, the rescaled no-jet case does a rather poor job of following the jet case. This may be because \citet{2015MNRAS.450.4035F} does a MHD calculation rather than a HD calculation as in \citet{2014ApJ...790..128F}.
    }
    \label{fig:fed_comp}
\end{figure*}

We have shown that $M_*(t) \propto (t-t_*)^2$ both with and without protostellar jet feedback, but that the stellar mass accreted after a given time is smaller by about a factor of 2.5 when jet feedback is included.
This result is also seen in \citet{2010ApJ...709...27W, 2014ApJ...790..128F, 2015MNRAS.450.4035F}.
While these authors did not note the power law dependence of mass upon time, it is clearly seen in their Figures 1, 9, and 2, respectively.
From these figures we have calculated the ratio of stellar mass in the no jet to jet case, finding the ratios to be 2.3, 2.4, and 3, respectively, which is similar to our ratio of 2.5.

The scaling between the jets and no-jet case of $f^2 = \left(1-f_{\rm jet}\right)^2$ that we have found above in equation (\ref{eq:mass_ratio}) appears to hold for \citet{2014ApJ...790..128F}.  This is shown in the left plot of Figure \ref{fig:fed_comp}.
Here we plot the SFE as a function of time from Figure 9 of \citet{2014ApJ...790..128F} for the jet (solid line) and no-jet (dashed line) cases.
We have also applied the simple rescaling $f^2=(1-f_{\rm jet})^2$ following equation (\ref{eq:mass_ratio}) where $f_{\rm jet} = 0.3$ in \citet{2014ApJ...790..128F}, the result is shown as the dotted line in Figure \ref{fig:fed_comp}.
The rescaled no-jet case does a surprisingly good job of following the jet case up to about an SFE of 0.4, which corresponds to a no-jet SFE of 0.8, after which the SFE of the no-jet case (unsurprisingly) turns over as there is little gas remaining to be accreted.
In the right plot of Figure \ref{fig:fed_comp}, we plot SFE as a function of time from Figure 2 of \citet{2015MNRAS.450.4035F} for the jet (solid line), no-jet (dashed line), and rescaled (dotted line) cases.
Here the rescaled no-jet case does a rather poor job of following the jet case.
This may be because \citet{2014ApJ...790..128F} (right plot) does a MHD calculation rather than a HD calculation as in \citet{2014ApJ...790..128F} (left plot).
 We do not believe that our analytic model can explain the evolution of $\dot{M}$ beyond $t/t_{ff} > 1.3$, nor do we think it should.
After a single free fall time $\approx 10\%$ of the entire gas mass in the simulation has been eaten by the created sink particles. By $t \approx 1.3 t_{ff}$ that has expanded to nearly $40\%$ of all the gas in the simulation. 
Once the mass of the stars becomes an appreciable fraction of the gas mass, accretion begins to be suppressed simply because there is insufficient gas.  This should occur when the mass in stars is a few tens of percent of the gas mass.

The scaling of equation (\ref{eq:mass_ratio}) may hold for other cases in the literature but we are unable to check them.
For example, it is unclear in \citet{2010ApJ...709...27W} what $f_{\rm jet}$ is equal to.
\citet{2014MNRAS.439.3420M} see the same curvature in $\dot{M}_*(t)$ that we do (their Figure 7), but they do not plot a no protostellar feedback simulation to compare against.

It appears that for hydrodynamic jet-feedback, the effect is to reduce the mass accretion rate by the rate at which the jet ejects material from the star (and disk).
Beyond this, the effects of hydrodynamic jet feedback appear to be minor.
The is clearly seen in the simulations of this paper and that of \citet{2014ApJ...790..128F}.
It may also be the case for other simulations in the literature, but such comparisons were not possible in those cases.

It does appear that magnetic field may enhance the effects of jet feedback.  The systematic evidence for this is sparse, but we can point to the comparison between the rescaled no-jet case and jet case of \citet{2015MNRAS.450.4035F} in the right plot of Figure \ref{fig:fed_comp}, where the rescaled case does not capture the complete effect of the jet.  Here the dynamics of the jet on the accreting gas appears to be more pronounced.  We should note, however, that the curvature in the SFE remains and suggests that some of the analytic results of MC15 may continue to hold in the MHD case.

\section{Conclusions}\label{sec:conclusions}
We performed simulations of turbulent, self-gravitating gas including star particle formation and protostellar jets. 
Starteing with uniform density in a box with length $16$ parsecs on a side, we drove turbulence until we reached a statistically steady state.
At that point, the density was no longer uniform.
We then turned on gravity and star formation.
We used AMR to follow collapsing regions down to an effective resolution of $32 \, K^3$ which gave us a $\Delta x$ of $100 {\rm AU}$ at the finest level of refinement.

We observed that the inclusion of protostellar jets does not affect the general dynamics of accreting gas.
In particular we saw $M_*(t) \propto f^2 (t-t_*)^2$ where $f = 1 - f_{\rm jet}$ is the fraction of mass accreted onto the protostar and $f_{\rm jet}$ is the fraction ejected by the jet.
We find that this mass ejection accounts for 75\% of the effect of jets on the star formation rate in our simulations.
This appears to be the case in similar simulations performed by other groups (e.g. see Figure \ref{fig:fed_comp}), but we find suggestions that this may be altered if MHD is included.

As we have found previously in the case without jets \citep{2017MNRAS.465.1316M}, the spherical average profile of gas around the protostar follows the analytic model of MC15 and does not seem the change in the case with jets.  In particular, the run of density finds an attractor solution prior to star formation and remains on that solution even after jets begin to blow out cavities in the surrounding medium.
The behavior of the infall and rotational velocities is similar regardless of whether jets are included or not.
The profile of the random velocities are also similar once the jet bi-cone is removed.  Finally, the mass accretion rates are similar in the jet and no-jet cases.

We also find that the collapse is outside-in  \citep{2017MNRAS.465.1316M}, and holds for both the jet and no jet simulations.
The average jet momentum per stellar mass does increase over time, though this is to be expected as the stars continue to accrete mass.
We did not run long enough for the stars to completely consume the surrounding gas and thus for the jets to begin to be shut off.
We find that jets do drive turbulence in the surrounding gas, but is confined to small scales of roughly a parsec.

\section*{Acknowledgements}

We thank Norm Murray for useful discussions and encouragement during the course of this work. We thank the anonymous referee for his/her detailed and thoughtful comments.
DM and PC are supported in part by the NASA ATP
program through NASA grant NNX13AH43G, NSF grant AST-1255469, and the University of Wisconsin-Milwaukee.
SG acknowledges support from  Summer Funded Internship Award at Skidmore College and Student Experiential Learning Fund at Dartmouth College.
Some of the computations were performed on
the gpc supercomputer at the SciNet HPC Consortium
\citep{2010JPhCS.256a2026L}. SciNet is funded by: the Canada
Foundation for Innovation under the auspices of Compute Canada; the
Government of Ontario; Ontario Research Fund - Research Excellence;
and the University of Toronto.
The authors also acknowledge the Texas Advanced Computing Center (TACC) at The University of Texas
at Austin for providing HPC resources that have contributed to the research results reported
within this paper. URL: \url{http://www.tacc.utexas.edu}




\bibliographystyle{mnras}
\bibliography{references} 




\appendix
\section{Effects of excising bi-cones aligned with protostellar jets on the infall, random, and rotational velocity} \label{sec:jet_cone_sub}
In this appendix we illustrate the effect of excising bi-cones aligned with the jet emitted by a protostar when calculating $\lvert u_r \rvert$, $v_{\rm T}$ and $v_\phi$. The simulated jets are powerful enough that they enforce outflow over almost the entire bi-cone over which the star particle emits the jet. However, because our jets are emitted along the instantaneous spin axis of the star particle, which is usually roughly perpendicular to the accretion disk, the jet outflow tends to avoid high density and infalling gas, which is generally near the plane of the disk. An example can be seen in Figure \ref{fig:jet_close_up}.

While the jets do not strongly affect the infall and rotational velocity of the bulk of the gas near the star particle, they do have a fairly strong effect on the random velocity, when averaged over spherical shells. A glance at Figure \ref{fig:jet_close_up} shows why: the jet velocities are very large in the evacuated region around the jet axis. Our calculation of $v_T$ involves subtracting the mass-weighted infall and rotational velocity in spherical shells from the velocity of each cell in the shell; since the jet expels low density gas, it does not affect the mass weighted infall or rotational velocity when performing the average, but it does boost the random velocity.

\begin{figure}
	\includegraphics[width=\columnwidth]{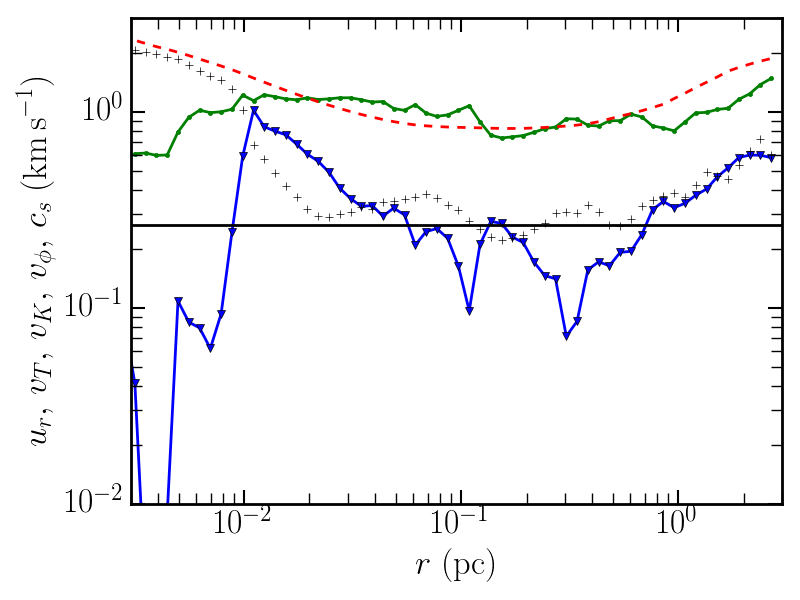}\\
    \includegraphics[width=\columnwidth]{velocity_ID4_shellsphere_124.png}\\
    \includegraphics[width=\columnwidth]{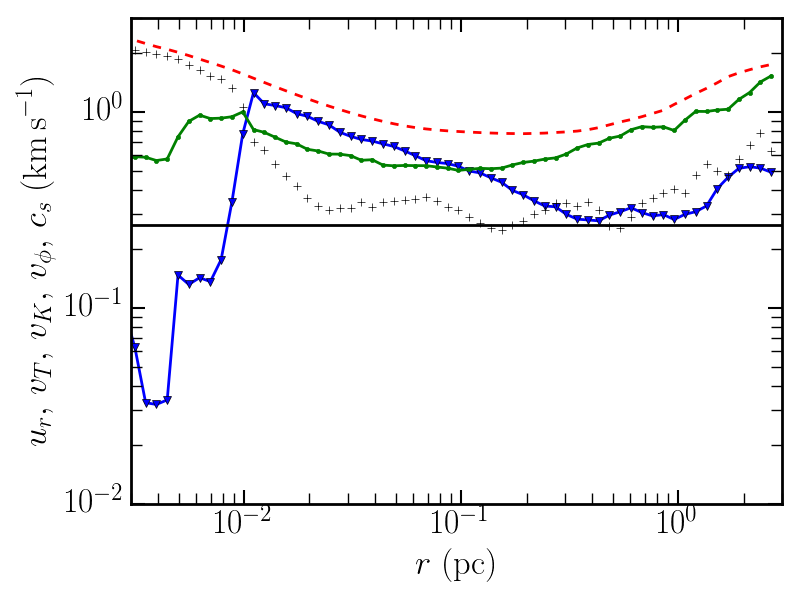}
    \caption{The run of velocity for a $M_*\approx 5.2 {\rm M_\odot}$ sink particle, averaged over spherical shells (top), averaged over the same shells but excising a bi-cone aligned with and having the same opening angle as the jet (middle) and excising a bi-cone with twice the jet opening angle (bottom). The sound speed is denoted by the black horizontal line while other lines denote the infall velocity $ \lvert u_r \rvert$ (blue triangles connected by a solid blue line), the random velocity $v_{\rm T}$ (green circles connected by a solid green line) and the rotational velocity $v_\phi$ (black crosses); the dashed red line shows $v_K \equiv \sqrt{GM(r)/r}$. Note that both $\lvert u_r \rvert$ and $v_{\rm T}$ increase with decreasing radius for $10^{-2}{\rm\, pc}<r<10^{-1}{\rm \, pc}$. }
    \label{fig:ideal_velocity}
\end{figure}

In Figure \ref{fig:ideal_velocity} we show three panels.
All three panels follow the same convention: the infall velocity $ \lvert u_r \rvert$ is depicted by the blue triangles, connected by a solid blue line,
the green circles connected by a solid green line show $v_{\rm T}$ while the black crosses show the rotational velocity $v_\phi$. The dashed red line shows $v_K \equiv \sqrt{GM(r)/r}$.
Finally, the sound speed is denoted by the black horizontal line.

All three snapshots are of the same particle, of roughly $5.2 {\rm M_\odot}$, at the same time, from the same simulation.
In the top panel we calculate the various velocities averaged over spherical shells, i.e.,  we do not remove the jet bi-cone, in the centre panel we remove the jet bi-cone, and in the bottom panel we remove a bi-cone with twice the jet opening angle.

The most dramatic change is in the turbulent velocity.
When we average over full spherical shells (top panel), the turbulent velocity remains roughly constant at all radii, with a slight increase at $r>1$ pc.
In the two plots where we excise the jet bi-cone (or a bi-cone with twice the jet opening angle), we see similar behavior to runs where no jets were included; at large radii the random velocity $v_{\rm T}$ decreases (by a factor of two) with decreasing radius, while at small radii ($r\lesssim 0.1$ pc) $v_{\rm T}$ increases with decreasing radius.

The effect of removing or not removing the jet bi-cone on $|u_r|$ and $v_\phi$ is much smaller.
The infall velocity smooths out slightly when we excise the region around the jet, but the general dynamic of decreasing velocity with decreasing radius at large $r$, and then inverting to increasing velocity with decreasing radius for $r\lesssim 0.1$ pc is seen in all three panels.

%
\section{Protostellar Model}
\label{appendix:protostar}
There is one final thing to note about the protostellar model presented by O09, which is the transition from no burning to core burning at a fixed temperature of $1.5 \times 10^6$ Kelvin.
When this transition is made, the polytropic index $n$ is changed from whatever its' current value is to $1.5$.
However, in changing $n$ we change the internal temperature of the protostar to a lower value than the required fixed temperature.
This raises the question of whether or not this has any effect on the radius of the star, the protostar's final mass, radius, and the length of time, for its evolution.
We did run a quick experiment with looping $n$ back and forth to achieve an actual burn temperature of $1.5 \times 10^6$ (rather than O09's dropping to a constant lower temperature).
What we found was that the end result sees no appreciable difference: roughly the same mass, same radius, and in roughly the same amount of time. The only modification was a change of $\approx 1/16 $th the radius during the core burning phase of the evolution of the protostar.

\bsp	
\label{lastpage}
\end{document}